%% file: main.tex
\documentclass[sigconf]{acmart}

\AtBeginDocument{%
  }

\copyrightyear{2024}
\acmYear{2024}
\setcopyright{acmlicensed}\acmConference[CCS '24]{Proceedings of the 2024 ACM SIGSAC Conference on Computer and Communications Security}{October 14--18, 2024}{Salt Lake City, UT, USA}
\acmBooktitle{Proceedings of the 2024 ACM SIGSAC Conference on Computer and Communications Security (CCS '24), October 14--18, 2024, Salt Lake City, UT, USA}
\acmDOI{10.1145/3658644.3690262}
\acmISBN{979-8-4007-0636-3/24/10}

\usepackage{tikz}
\usepackage{amsmath}
\usepackage{graphicx}

\usepackage{pgfplots}
\pgfplotsset{compat=1.9}

\usepackage{listings}
\usepackage{upquote}

\usepackage[htt]{hyphenat}
\usepackage{enumerate}
\usepackage{enumitem}
\usepackage{graphicx}
\usepackage{epstopdf}
\usepackage{epsfig}
\usepackage{url}
\usepackage{etoolbox}
\usepackage{tikz}
\usepackage{amsmath}
\usepackage{stmaryrd}
\usepackage{mathtools}
\usepackage{hhline}
\usepackage{setspace}
\usepackage{booktabs}
\usepackage{multicol}
\usepackage{multirow}
\usepackage{xcolor,colortbl}
\usepackage{array}
\usepackage{xspace}
\usepackage{pifont}
\usepackage{csquotes}
\usepackage{subcaption}
\usepackage[flushleft]{threeparttable}

\definecolor{lightgray}{rgb}{0.95, 0.95, 0.95}
\definecolor{darkgray}{rgb}{0.4, 0.4, 0.4}
\definecolor{editorGray}{rgb}{0.95, 0.95, 0.95}
\definecolor{editorOcher}{rgb}{1, 0.5, 0} % #FF7F00 -> rgb(239, 169, 0)
\definecolor{editorGreen}{rgb}{0, 0.5, 0} % #007C00 -> rgb(0, 124, 0)
\definecolor{orange}{rgb}{1,0.45,0.13}		
\definecolor{olive}{rgb}{0.17,0.59,0.20}
\definecolor{brown}{rgb}{0.69,0.31,0.31}
\definecolor{purple}{rgb}{0.38,0.18,0.81}
\definecolor{lightblue}{rgb}{0.1,0.57,0.7}
\definecolor{lightred}{rgb}{1,0.4,0.5}
\definecolor{azure}{RGB}{0,127,255}
\definecolor{Gray}{gray}{0.9}
\definecolor{gray1}{rgb}{193,199,198}
\definecolor{editorGray}{cmyk}{0, 0, 0, 0.2}
\definecolor{editorOcher}{rgb}{1, 0.5, 0}
\definecolor{eclipseStrings}{RGB}{42,0.0,255}
\definecolor{eclipseKeywords}{RGB}{127,0,85}
\colorlet{numb}{magenta!60!black}
\lstdefinelanguage{json}{
  keywordstyle=\color{red}\bfseries,
 keywords={},
  keywordstyle=\color{red},
  identifierstyle=\color{black},
  sensitive=false,
      breaklines=true,
    frameround=ffff,
    frame=single,
  comment=[l]{//},
  morecomment=[s]{/*}{*/},
  commentstyle=\color{purple}\ttfamily,
  stringstyle=\color{red}\ttfamily,
    literate=
     *{0}{{{\color{numb}0}}}{1}
      {1}{{{\color{numb}1}}}{1}
      {2}{{{\color{numb}2}}}{1}
      {3}{{{\color{numb}3}}}{1}
      {4}{{{\color{numb}4}}}{1}
      {5}{{{\color{numb}5}}}{1}
      {6}{{{\color{numb}6}}}{1}
      {7}{{{\color{numb}7}}}{1}
      {8}{{{\color{numb}8}}}{1}
      {9}{{{\color{numb}9}}}{1}
      {\{}{{{\color{delim}{\{}}}}{1}
      {\}}{{{\color{delim}{\}}}}}{1}
      {[}{{{\color{delim}{[}}}}{1}
      {]}{{{\color{delim}{]}}}}{1},
}

\definecolor{delim}{RGB}{20,105,176}
\definecolor{numb}{RGB}{106, 109, 32}
\definecolor{string}{rgb}{0.64,0.08,0.08}
\definecolor{light-gray}{gray}{0.95}

\lstdefinelanguage{JavaScript}{
  keywords={typeof, new, true, false, catch, function, return, null, catch, switch, var, if, in, while, do, else, case, break, for, forEach},
  keywordstyle=\color{blue}\bfseries,
  ndkeywords={class, export, boolean, throw, implements, import, this, name},
  ndkeywordstyle=\color{darkgray}\bfseries,
  identifierstyle=\color{black},
  sensitive=false,
  comment=[l]{//},
  morecomment=[s]{/*}{*/},
  commentstyle=\color{purple}\ttfamily,
  stringstyle=\color{red}\ttfamily,
  morestring=[b]',
  morestring=[b]"
}

\lstdefinelanguage{HTML5}{
        language=html,
        sensitive=true, 
        alsoletter={<>=-},
        otherkeywords={
        <html>, <head>, <title>, </title>, <meta, />, </head>, <body>,
        <canvas, \/canvas>, <script>, </script>, </body>, </html>, <!, html>, <style>, </style>, ><
        },  
        ndkeywords={
        =,
        charset=, id=, width=, height=,
        border:, transform:, -moz-transform:, transition-duration:, transition-property:, transition-timing-function:
        },  
        morecomment=[s]{<!--}{-->},
        tag=[s]
}

\usepackage{algorithm}
\usepackage{algpseudocode}

\algnewcommand\algorithmicforeach{\textbf{for each}}
\algdef{S}[FOR]{ForEach}[1]{\algorithmicforeach\ #1\ \algorithmicdo}

\usepackage{booktabs}

\usepackage{xspace}

\newcommand{\tikzcircle}[2][red,fill=red]{\tikz[baseline=-0.5ex]\draw[#1,radius=#2] (0,0) circle ;}%

\definecolor{ForestGreen}{RGB}{34,139,34}

\newcommand{\js}{JavaScript\xspace}
\newcommand{\systemLong}{Unbundle-Rewrite-Rebundle\xspace}
\newcommand{\system}{URR\xspace}
\newcommand{\ie}{i.e.,}
\newcommand{\eg}{e.g.,}

\begin{document}

%%
%% The "title" command has an optional parameter,
%% allowing the author to define a "short title" to be used in page headers.
\title[\systemLong]{\systemLong: Runtime Detection and Rewriting of Privacy-Harming Code in \js Bundles}

%%
%% The abstract is a short summary of the work to be presented in the
%% article.
\input{00-abstract}

\author{Mir Masood Ali}
% \orcid{0009-0006-8462-0945}
\affiliation{
\institution{University of Illinois Chicago}
\city{Chicago}
\country{USA}}
% \email{mali92@uic.edu}

\author{Peter Snyder}
% \orcid{0000-0001-7880-2503}
\affiliation{
\institution{Brave Software}
\city{San Francisco}
\country{USA}}
% \email{pes@brave.com}

\author{Chris Kanich}
% \orcid{0000-0002-3836-2168}
\affiliation{
\institution{University of Illinois Chicago}
\city{Chicago}
\country{USA}}
% \email{ckanich@uic.edu}

\author{Hamed Haddadi}
% \orcid{0000-0002-5895-8903}
\affiliation{
\institution{Imperial College London \& Brave Software}
\city{London}
\country{UK}}
% \email{h.haddadi@imperial.ac.uk}

%%
%% The code below is generated by the tool at http://dl.acm.org/ccs.cfm.
%% Please copy and paste the code instead of the example below.
%%
\begin{CCSXML}
<ccs2012>
   <concept>
       <concept_id>10002978.10003022.10003026</concept_id>
       <concept_desc>Security and privacy~Web application security</concept_desc>
       <concept_significance>500</concept_significance>
       </concept>
 </ccs2012>
\end{CCSXML}

\ccsdesc[500]{Security and privacy~Web application security}

%%
%% Keywords. The author(s) should pick words that accurately describe
%% the work being presented. Separate the keywords with commas.
\keywords{Web privacy; Content blocking}

\maketitle

\input{01-alt_intro}
\input{02-background}

\input{03-framework}
\input{04-evaluation}
\input{05-discussion}

\input{06-relatedwork}
\input{07-conclusions}

\section*{Acknowledgements}

We thank Victor Escudero, Marek Cwiek, and Zaheer Safi for performing the manual website analysis. 
% We also thank the anonymous reviewers for their helpful feedback. 
This material is based upon work supported by the National Science Foundation (CNS-2247515). 
% The views in this paper are only those of the authors and may not reflect those of the US Government or the NSF.

% \section*{Availability}

% Artifacts are available at \url{https://github.com/masood/urr}.

%%
%% The next two lines define the bibliography style to be used, and
%% the bibliography file.
\bibliographystyle{ACM-Reference-Format}
\bibliography{main}

%%
%% If your work has an appendix, this is the place to put it.
% \appendix
% \input{08-artifact}

\end{document}

%% file: 00-abstract.tex
%-------------------------------------------------------------------------------
\begin{abstract}
%-------------------------------------------------------------------------------

This work presents \systemLong{} (\system), a system for detecting privacy-harming portions of bundled \js{} code and rewriting that code \emph{at runtime} to remove the privacy-harming behavior without breaking the surrounding code or overall application. \system{} is a novel solution to the problem of \js{} bundles, where websites pre-compile multiple code units into a single file, making it impossible for content filters and ad-blockers to differentiate between desired and unwanted resources. Where traditional content filtering tools rely on URLs, \system{} analyzes the code at the AST level, and replaces harmful AST sub-trees with privacy-and-functionality maintaining alternatives.

We present an open-sourced implementation of \system{} as a Firefox extension and evaluate it against \js{} bundles generated by the most popular bundling system (Webpack) deployed on the Tranco 10k. We evaluate \system by precision (1.00), recall (0.95), and speed (0.43s per script) when detecting and rewriting three representative privacy-harming libraries often included in \js{} bundles, and find \system{} to be an effective approach to a large-and-growing blind spot unaddressed by current privacy tools.

\end{abstract}

%% file: 01-alt_intro.tex
\section{Introduction}
An enormous body of research has established Web content filtering (e.g.,
blocking advertising, tracking, and other unwanted network requests on websites)
as an important and effective technique for improving privacy\cite{merzdovnik2017block, gervais2017quantifying}, security\cite{li2012knowing, zarras2014dark}, and performance\cite{garimella2017ad, pujol2015annoyed}. Most Web content filtering
approaches rely on crowd-sourced lists of regular-expression-like rules that
describe which URLs the browser should load and which should be blocked.

This approach---broadly, URL-based content filtering---works because
URLs, in practice, provide useful and stable information about the resources they
map to. In some cases, this is because of the text in the URL (e.g., browsers
can make a reasonable guess about the purpose of JavaScript returned from
a URL like
(\texttt{https://advertising\\.example/tracker.js}), or because experts
have manually evaluated the resource returned from a URL and found it to
be similarly harmful to users.

However, modern Web development practices make URL-based content filtering increasingly difficult. Previously, Web applications were often delivered as a collection of discrete \js{} files, each fetched independently from their own URL (\eg{} \texttt{/script/library.js}, \texttt{/script/tracker.js},\texttt{/script/app.js}), which allowed URL-based content filtering tools to easily block some parts of an application, but not others. Increasingly, though, developers integrate bundling tools as part of their build and deployment practices, compiling all of the libraries and application code into a single file unit, which is delivered to the browser from a single URL (\eg{} \texttt{/script/bundle.js}). 

These bundling approaches, inadvertently or otherwise, circumvent
URL-based content filtering tools. When applications are delivered as a single bundled code unit, URL-based filtering tools
can no longer block just parts of the application; blocking a site's \js{} becomes an all-or-nothing proposition. And since blocking all \js{} on a page breaks useful functionality on many sites, in practice, content filtering tools are reduced to blocking nothing, reintroducing the privacy, security, and performance issues the user wanted to avoid in the first place.
% by combining multiple individual JavaScript code units into a single ``bundled''
% code unit. Where sites might have previously included their Web application
% as discrete, independently fetched JavaScript files, application developers
% increasingly fetch each required JavaScript library at application build time,
% package them into a single resource, and instead have users download and execute
% a single, combined JavaScript file. In other words, web applications that
% previously fetched three JavaScript files
% (e.g., \texttt{/script/library.js}, \texttt{/script/tracker.js}, and
% \texttt{/script/app.js}), are increasingly deployed by combining all
% referenced code units into a single JavaScript resource (e.g., \texttt{/script/bundle.js}).

% There is a wide range of bundling techniques used on the Web. The most
% common kinds of JavaScript bundling are tools like
% Webpack\footnote{\url{https://webpack.js.org/}}, which bundle multiple JavaScript
% libraries into a single sources as part of a Node based build process.
% Google has proposed a similar approach called
% Web Bundles\footnote{\url{https://developer.chrome.com/docs/web-platform/web-bundles/}},
% which allow entire Web pages and their sub-resources (e.g., JavaScript, CSS, images)
% to be bundled into a single archive. And CDN companies like Cloudflare
% offer systems like ``Managed Components''\footnote{\url{https://managedcomponents.dev/}}
% that allow bundling-as-a-service at the network edge.

This work presents the design and implementation of \systemLong{} (\system), a system to enable
content blocking in modern Web applications, even when Web applications
are deployed as a compiled, single file \js{} bundle. In other words, \system{} aims
to enable browsers to avoid executing the code from \texttt{/script/tracker.js},
while still executing the
non-privacy harming code originally provided in \texttt{/script/library.js} and
\texttt{/script/app.js}, \emph{even when} all three libraries are bundled and
delivered in \texttt{/script/bundle.js}. % a single Web bundle.

\system{} is a novel, practical solution to a problem that a vein of related Web privacy research has explored. Works like~\cite{trackersift} identify
that bundled applications are widespread and pose a serious challenge to Web privacy,
and \cite{smith2022blocked} found that blocking these bundled \js{} resources often
broke the benign, desirable parts of websites. Systems like~\cite{SugarCoat:2021}
showed that bundled applications could be automatically rewritten to prevent
privacy harm, though with expensive precomputation, which rendered practical deployment
prohibitive. \system{} is a first-in-class approach to solving the privacy and security
harms caused by bundled \js{} applications in a performant and practical method.

To do so, \system{}  solves several non-trivial challenges:

\textbf{First}, the system must identify known privacy-harming code (e.g., the code 
delivered from \texttt{/script/tracker.js}) within the larger bundled application, without any information about the URL from which the code originally came. This identification must be robust even across the kinds of code modifications and transformations \js{} bundlers make in their build processes (\eg{} minification, dead-code elimination, tree-shaking).

% This problem is made even more difficult because popular bundling systems
% post-process, minify, ``tree-shake'', and otherwise optimize each library
% during the bundling process. As a result, code may change significantly
% in both presentation (i.e., labels) and structure (i.e., top level functions,
% classes, constants, etc) during the bundling process, making reidentification
% difficult.

\textbf{Second}, \system{} must remove known-privacy-harming code from the bundled 
application without breaking desirable functionality in the surrounding code. 
Just deleting unwanted libraries
from a bundled application will, in practice, be counter-productive since the
application will fail when trying to access now-deleted functions and classes
defined by the deleted library. A useful solution must remove the unwanted,
target libraries from the bundled application \emph{without breaking surrounding
code}.

\textbf{Third}, such a system must be performant and be able to Unbundle,
analyze, modify, and reconstitute bundled Web applications at runtime, and quickly,
in a way that maintains the usefulness of the Web application. If the performance
overhead of a privacy-and-security preserving system is too costly, then the system
is, in practice, unusable and so not meaningfully useful to benefit users.

\system{} is implemented in several parts: i. as a database of signatures of
ASTs of real-world known-privacy-harming code, ii. a library of crowd-sourced
privacy-preserving alternative implementations of privacy-harming code designed to remove
privacy harming behaviors without impacting the surrounding application code, iii.
a browser extension that, at runtime, decomposes a bundled \js{} application
into its constituent libraries, detects the sub-ASTs from the bundled application
that come from known-privacy-harming libraries and rewrite the bundled application
with the stub libraries in place of the privacy-harming versions.

% This paper presents \system, a system dynamically blocking known-harmful JavaScript
% code from bundled Web applications, without harming user-beneficial behaviors
% in those Web applications. \system works by:
% \kaytwo{this feels out of place: how we do it should either not be in the intro or should only be very succinctly summarized, more important to highlight contributions)}
% \begin{itemize}[leftmargin=*,nosep,noitemsep]
%   \item offline building a database of signatures of the ASTs of top level
%     structures in targeted (i.e., unwanted) JavasScript libraries
%   \item modifying a commodity Web browser (via a Firefox browser extension) to
%     identify and unbundle Web applications into their component modules/sub-libraries,
%   \item searching for known-unwanted libraries in bundled
%     application by generating signatures for each sub-module, and checking
%     them against the offline generated database of known-unwanted code
%   \item if a match is found, then replacing the target code with a pre-authored
%     compatibility-maintaining shim implementation of the target library,
%     and then passing the the now-modified bundled application to the
%     browser's JavaScript engine.
% \end{itemize}

% \subsection{Contributions}
More broadly, this work makes the following contributions to Web privacy and security.

\begin{enumerate}[leftmargin=*,nosep,noitemsep]
  \item A novel algorithm for efficiently generating fingerprints of
    bundled modules within JavaScript libraries. These fingerprints are
    robust to many of the modifications that bundling tools make during
    their complication process (e.g., label minification, ``tree-shaking'').
  \item An open-sourced, empirically tuned system for:
    \begin{enumerate}
      \item unbundling applications into their constituent libraries
      \item generating fingerprints for each sub-library and checking
        them against a database of known unwanted JavaScript libraries
      \item replacing unwanted JavaScript libraries with compatibility-preserving
        ``shim'' implementations, which maintain the library's API ``shape'',
        while removing any privacy-or-security affecting behaviors
      \item reconstituting the resulting new application into a new bundle that
        can then be passed to the browser's JavaScript engine for normal
        execution.
    \end{enumerate}
  \item An empirical evaluation of the accuracy and performance of our
    system when applied to a representative crawl of the Web, finding
    that our system results in libraries being blocked on 7\% of the top 10K sites in the Tranco list within practical performance bounds.
  \item An open-source implementation of our system as a Firefox extension,
    along with the complete dataset for all discussed figures and measurements, available at \textcolor{blue}{\url{https://github.com/masood/urr}}. 
    % We have uploaded artifacts to HotCRP along with this submission.
  \end{enumerate}

%% file: 02-background.tex
%-------------------------------------------------------------------------------
\section{Background}
\label{sec:background}
%-------------------------------------------------------------------------------

This section first provides a primer on concepts relevant to bundling and their use in web development. It then presents a simple example that highlights the limitations of existing content-blocking approaches. The section concludes by outlining the properties of an effective solution.

\subsection{Bundles and Relevant Concepts}
\label{sec:bundles-background}

As websites and web applications grow more complex, they require a plethora of functionality, often aided by numerous libraries and dependencies. Handling these dependencies can prove to be quite strenuous, especially when considering the range of platforms on which the code needs to execute correctly.
\js bundlers are essential tools used by web developers to streamline the handling of code and dependencies within complex web applications. At its core, a \js bundler is a utility that gathers and wraps code from multiple \js files.
% 
% Consider the example shown in Figure~\ref{fig:motivating-example}. Initially (left half), the website spread its code across three \texttt{<script>} tags. After bundling its resources (right half), the website included and executed both, first-party and third-party code, with a single call to a bundled resource. 
% 
Bundles not only reduce the number of network requests required to load a web page but also optimize the handling of dependencies and various aspects of development and production environments.
The simplicity of the example hides certain nuances necessary to understand the complexity of bundles. Below, we introduce a few fundamental concepts that can help better understand \js bundles.

\subsubsection{Modular Programming.} Web developers design and create websites in different environments than browsers. These environments have their own caveats and follow different programming concepts and philosophies. Modular programming is a general programming concept where developers separate complex application code into independent pieces called \textit{modules}. A module forms the atomic unit of a bundle and roughly corresponds to a code snippet relevant to a file or library that exports functionality that is consumed by other modules. Popular \js development environments like Node.js\footnote{\url{https://nodejs.org/en/}} adopt a modular programming approach. Developers can create their own modules and use reuse modules available in the \texttt{npm} registry\footnote{https://www.npmjs.com/}.

\subsubsection{\js Module Systems.} Like development environments and programming approaches, \js module systems are also not a monolith. Depending on the context in which they are consumed and executed, \js modules express functionality in different ways. The two most popular module systems for \js are (1) the ECMAScript modules (ESM), which are consumed with \texttt{import} statements, and (2) CommonJS (CJS) modules, which are consumed with \texttt{require} statements. While Node.js supports both types of modules, web browsers only recognize \texttt{import} statements. Bundles, therefore, include wrappers around modules and provide workarounds for \texttt{require} statements within web browsers. 

\subsubsection{Inter-module Dependencies.} Larger projects include multiple libraries and packages and, as a result, comprise numerous interdependent modules. When bundles gather and parse all the modules that need to be combined, they create a dependency graph that helps determine (1) the order and chain of dependencies between modules and (2) which code snippets can be combined within a module and which snippets need to be split across multiple modules.

\subsubsection{Minification.} Bundlers additionally perform a code transformation step that reduces the overall size of the code. Depending on various configuration and optimization options, minification returns an irreversible code output containing randomized, unidentifiable variable names and changes to the code syntax that only retains its underlying logic.

\subsubsection{Source Maps.} During development, bundles provide source maps as a key to reverse minification and debug parts of code. These files help map minified code back to their unminified counterparts and, hence, identify individual modules. However, source maps are unavailable by default in production, making it difficult to reverse engineer bundled code in the wild.

\subsubsection{Popular Bundlers.} Examples of popular \js bundlers include Webpack\footnote{\url{https://webpack.js.org/}}, Browserify\footnote{\url{https://browserify.org/}}, Rollup\footnote{\url{https://rollupjs.org/}}, and Parcel~\footnote{\url{https://parceljs.org/}}, each of which offer unique features and advantages. In this work, we focus on Webpack because it is the most popular and mature \js bundler as determined from GitHub stars\footnote{\url{https://github.com/webpack/webpack}}, NPM weekly downloads\footnote{\url{https://www.npmjs.com/package/webpack}}, and prior work~\cite{Rack2023CCS}.

\begin{figure}[t]
    \begin{lstlisting}[language=Javascript,firstnumber=1, caption=A non-minified example of a webpack bundle.,
            label={lst:example-bundled-code}]
(function (modules) {
    // The module cache
    var installedModules = {};
    
    // The require function
    function __webpack_require__(moduleId) {
        // Check if module is in cache
        if (installedModules[moduleId]) {
        return installedModules[moduleId].exports;
        }
        // Create a new module (and put it into the cache)
        var module = (installedModules[moduleId] = {
        exports: {},
        });
    
        // Execute the module function
        modules[moduleId].call(
        module.exports,
        module,
        module.exports,
        __webpack_require__
        );
    
        // Return the exports of the module
        return module.exports;
    }
    
    // Load entry module and return exports
    return __webpack_require__(0);
    })([
    /* 0 */
    function (module, exports, __webpack_require__) {
        const hello = __webpack_require__(1);
        console.log(hello.sayHello("Webpack"));
    },
    /* 1 */
    function (module, exports) {
        exports.sayHello = function (name) {
        return "Hello, " + name + "!";
        };
    },
]);              
\end{lstlisting}
\vspace{-4ex}
\end{figure}

\subsubsection{General Structure of Webpack Bundles.}

Listing~\ref{lst:example-bundled-code} presents an example of a script comprising a webpack bundle. We describe the example below.

\begin{itemize}[leftmargin=*,nosep,noitemsep]
    \item[\ding{228}] Webpack wraps the bundled code in an \textbf{Immediately Invoked Function Expression (IIFE)}. As a result, the bundle is executed as soon as it is loaded, thereby making all necessary functions and variables available in the global scope.
    \item[\ding{228}] The bundle comprises a modular system where each module is represented as a function in an \textbf{array or object of modules}. Listing~\ref{lst:example-bundled-code} contains a module array in L30-42.
    \item[\ding{228}] \textbf{Function wrappers around each module} handle dependencies on other modules and gather any variables and functions exported by the module. 
    \item[\ding{228}] The \texttt{\_\_webpack\_require\_\_} function (Listing~\ref{lst:example-bundled-code}, L6-26) loads and executes modules and \textbf{manages exports and dependencies}.
    \item[\ding{228}] The bundle executes an \textbf{entry point module} (module 0 in Listing~\ref{lst:example-bundled-code}). The entry point module uses \texttt{\_\_webpack\_require\_\_}  to load functionality from other modules (Listing~\ref{lst:example-bundled-code}, L33).
\end{itemize}

Overall, bundlers allow developers to organize their code into modules, manage dependencies, and apply various optimizations like minification and code splitting, resulting in more efficient, maintainable, and faster-loading web applications.

\subsection{Motivating Example}
\label{sec:motivating-example}

In this section, we present an example that shows how typical content-blocking approaches work and why they are ineffective against bundles.

% \begin{figure}[t]
% \begin{lstlisting}[language=HTML,firstnumber=1, caption=Example Code Inclusion.,
%         label={lst:example-code-inclusion}]
% <script src="/setup.js"></script>
% <script src="tracker.com/track.js"></script>
% <script>
%     setup();
%     track();
% </script>
% \end{lstlisting}
% \end{figure}

% \begin{figure}[t]
%     \begin{lstlisting}[language=HTML,firstnumber=1, caption=Example Bundle Inclusion.,
%             label={lst:example-bundle-inclusion}]
% <script src="bundle.js"></script>
% <script>
%     setup();
%     track();
% </script>
% \end{lstlisting}
% \end{figure}

% \begin{figure}[t]
%     \begin{lstlisting}[language=Javascript,firstnumber=1, caption=Example Bundled Code.,
%             label={lst:example-bundled-code}]
% wrapperObject = {
%     a123: function(e,t,n) {/** Setup Code */},
%     b456: function(e,t,n) {/** Track Code */}
% }
% handleWrappers = function() {
%     /** expose wrapperObject */
% }
% \end{lstlisting}
% \end{figure}

\begin{figure*}[ht]
    \centering
    \includegraphics[width=0.8\textwidth,keepaspectratio]{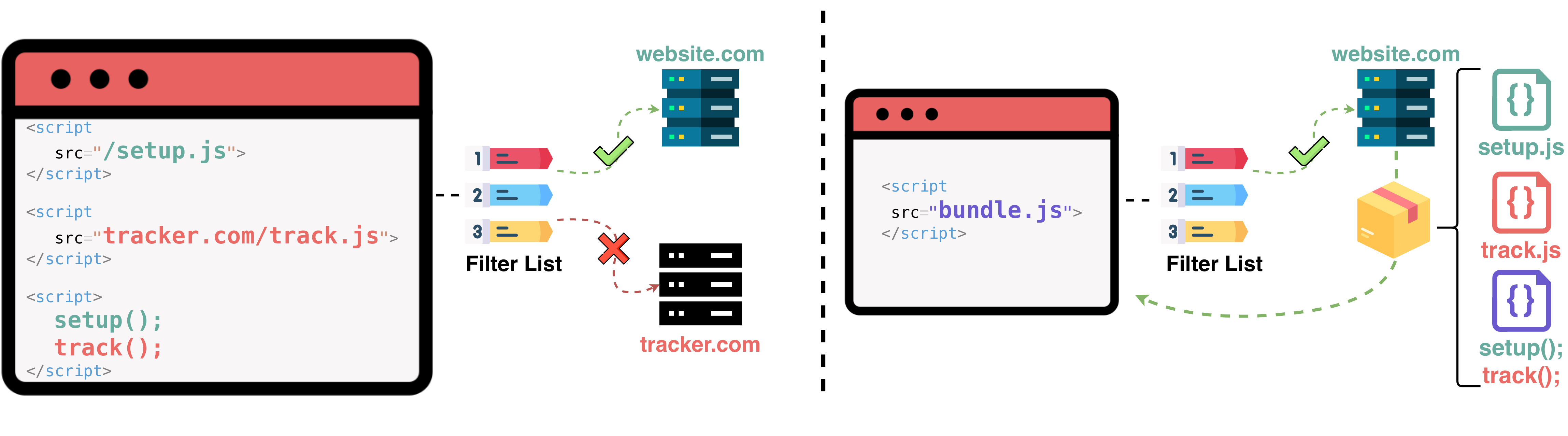}
    \caption{Motivating example.}
    \label{fig:motivating-example}
\end{figure*}

\subsubsection{Generic Code Inclusion.} Consider the toy example presented on the left half of Figure~\ref{fig:motivating-example}. Here, the website loads two scripts, each of which defines global functions. The first script, \texttt{setup.js}, loads from the website's domain itself and contains benign code relevant to the website's functionality. The second script, \texttt{track.js}, loads from a known tracking domain, \texttt{tracker.com}, and returns a script that creates a unique identifier to track the user. When the website includes the two scripts from two separate \texttt{<script>} tags, it triggers two network requests, one to \texttt{website.com} and another to \texttt{tracker.com}.

\subsubsection{Typical Content Blocking Approach.} A typical approach to blocking privacy-harming scripts involves the use of a curated list of domains and regular expressions (for example, from EasyList, EasyPrivacy\footnote{\url{https://easylist.to/}}). These lists are developed from manual contributions and include domains and paths to known privacy-harming resources. Content blocking tools (e.g., AdBlock, uBlock) pull from these filter lists. The tools intercept outgoing network requests, compare them against entries in filter lists, and create interventions if they find a match. In the toy example, a filter list contributor might add the rule \texttt{||tracker.com/track.js}. Thereafter, when the website creates two network requests, the content-blocking tool permits a request to \texttt{script.js} but blocks the request to \texttt{tracker.com/track.js}. This way, the privacy-harming script is neither loaded nor executed in the user's browsing session.

\subsubsection{Limitations of Existing Content Blocking Approaches.} Consider the scenario presented in the right half of Figure~\ref{fig:motivating-example}. Unlike the previous example, the website includes a single \texttt{<script>} tag that fetches code from its own server. The resulting network request does not have a corresponding entry in the filter list and is, therefore, not blocked by the content-blocking tool. The server responds with a \texttt{bundled} script that includes both benign code (\texttt{setup()}) and privacy-harming code (\texttt{track()}). Bundled resources expose multiple shortcomings of existing content-blocking approaches, which we briefly describe below. 
\begin{itemize}[leftmargin=*,nosep,noitemsep]
    \item[\ding{228}] Privacy-harming code can be fetched from \textbf{multiple, variable resource paths}, including from first-party domains, making it impossible for contributors to manually detect and curate a comprehensive, non-exhaustive list of entries.
    \item[\ding{228}] Bundles include code from multiple resources. As a result, privacy-harming code is embedded along with necessary and benign functionality. Blocking the network request itself can break the website. Existing approaches must adopt an approach that targets \textbf{embedded resources} instead of the entire resource itself.
    \item[\ding{228}] Bundles also \textbf{mutate code included in modules} with wrappers and through various minification and obfuscation techniques. These mutations make it difficult to identify content with comparisons against regular expressions.
\end{itemize}

\subsection{Properties of an Ideal Solution}
\label{sec:properties-of-an-ideal-solution}

We outline the properties of a general solution to identifying and replacing privacy-harming modules from bundled scripts.

First, a robust solution should be able to \textbf{identify and target specific modules}. Unlike typical content-blocking approaches, the solution cannot rely on domain-based blocking or attempt to replace the script entirely.

Second, as a corollary to the previous point, the solution should have \textbf{limited impact on benign functionality}, \ie{} the solution should limit the effect of its intervention to privacy-harming code, leaving execution of other modules untouched. This also involves ensuring that any dependencies on the targeted module are handled in a way that limits side effects.

Third, the solution should be \textbf{generalizable across multiple dimensions}. It should apply to multiple privacy-harming libraries (and multiple versions of libraries) that interest content-blocking tools.  Additionally, it should also be generalizable to different bundling strategies and robust against minification and obfuscation techniques.

Finally, the solution should have a \textbf{limited performance overhead}. While the solution executes in real-time, \ie{} as scripts are loaded and executed in the browser, it needs to limit its effect on the usability of websites.

% \begin{enumerate}
    % \item[\ding{228}] Can identify specific parts of bundled code
    % \item[\ding{228}] General enough to work across different libraries and bundling strategies
    % \item[\ding{228}] Limited performance overhead
    % \item[\ding{228}] Does not affect website functionality -- transarent to the user
% \end{enumerate}

%% file: 03-framework.tex
%-------------------------------------------------------------------------------
\section{\systemLong{} Design}
\label{sec:framework}
%-------------------------------------------------------------------------------

\begin{figure*}[ht]
    \centering
    \includegraphics[width=0.85\textwidth,keepaspectratio]{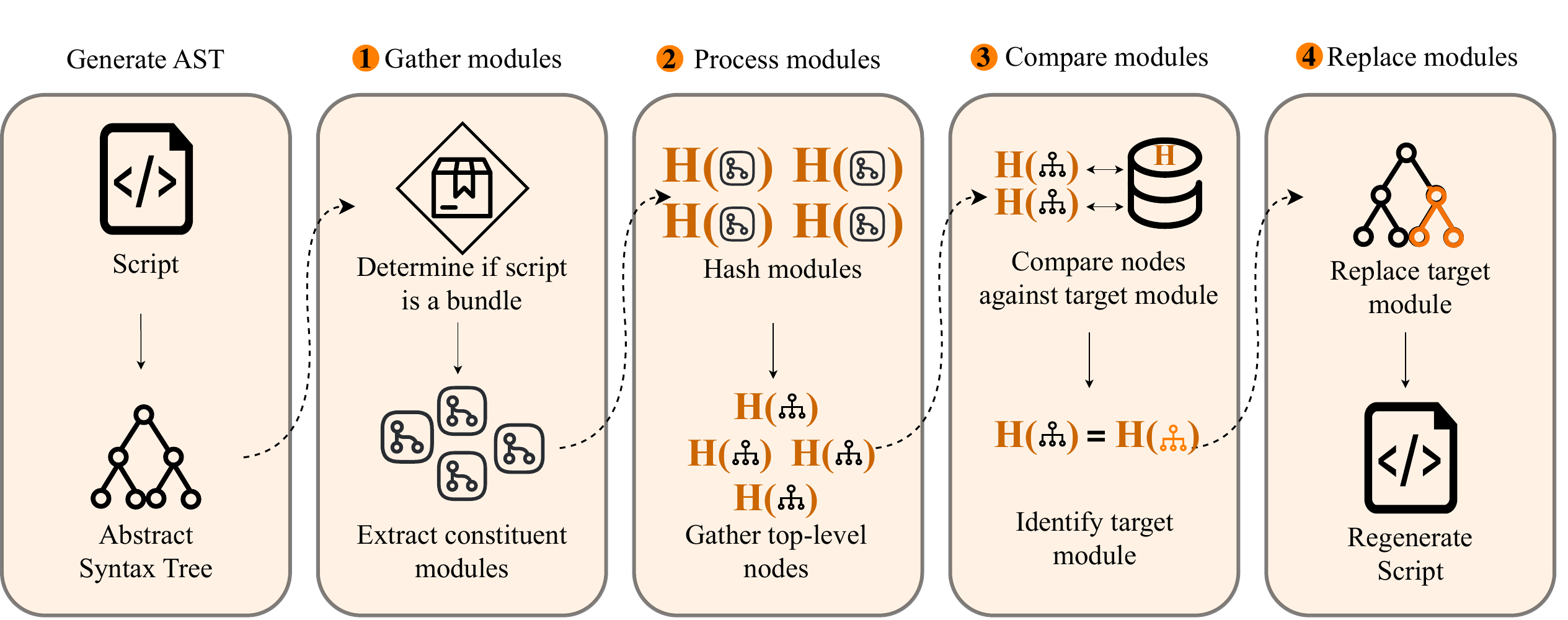}
    \caption{Overview of the framework.}
    \label{fig:framework-overview}
\end{figure*}

% In this section, we present the design of \system, a system that identifies that replaced privacy-invasive code within bundled scripts. 
\systemLong{} (\system) adopts a static analysis approach that leverages the code structures of privacy-harming modules to identify and neutralize them when embedded in bundled scripts without disrupting the functionality of other application components. \system adopts a four-step process (see Figure~\ref{fig:framework-overview}):
\begin{itemize}[leftmargin=*,nosep,noitemsep]
    \item[\ding{228}] First, \system generates an Abstract Syntax Tree (AST) representation of a given script. It then analyzes the structure to identify if the script is a bundle that comprises one or more modules. If so, \system gathers the component modules for further analysis.
    \item[\ding{228}] Next, \system processes each module into an implementation agnostic representation, i.e., stripping variable names, function names, and object properties. It creates a bottom-up hash of the AST structure and uses the resulting representation for comparison.
    \item[\ding{228}] Next, \system compares each processed module against previously generated representations of privacy-harming modules. If \system finds a match, it marks the corresponding module for replacement.
    \item[\ding{228}] Finally, \system replaces each marked privacy-harming module with a corresponding benign replacement. In doing so, \system ensures that access and use of the replacement does not break other parts of website functionality. It then stitches the bundled script and moves it along for consumption and execution.
\end{itemize}

All the steps mentioned above are performed in real-time, i.e. when resources are loaded in the user's browser. However, processed representations of target modules and benign replacements are created and gathered offline.

% Please add the following required packages to your document preamble:
% \usepackage{booktabs}
% \usepackage{multirow}
% \usepackage{graphicx}
% \usepackage[table,xcdraw]{xcolor}
% Beamer presentation requires \usepackage{colortbl} instead of \usepackage[table,xcdraw]{xcolor}
\begin{table*}[t]
    \centering
    \caption{Refining code to identify bundled webpack modules within scripts.}
    \label{tab:bundle-sampling}
    \resizebox{0.8\textwidth}{!}{%
    \begin{tabular}{@{}llrrlrrllrllr@{}}
    \toprule
     &
      \multicolumn{3}{c}{\textbf{Round 1 (n = 288)}} &
      \multicolumn{3}{c}{\textbf{Round 2 (n = 432)}} &
      \multicolumn{3}{c}{\textbf{Round 3 (n = 614)}} &
      \multicolumn{3}{c}{\textbf{Spot Check (n=300)}} \\ \midrule 
     &
      \multicolumn{1}{c}{\cellcolor[HTML]{FFFFFF}} &
      \multicolumn{2}{c}{Code} &
      \multicolumn{1}{c}{\cellcolor[HTML]{FFFFFF}} &
      \multicolumn{2}{c}{Code} &
       &
      \multicolumn{1}{c}{\cellcolor[HTML]{FFFFFF}} &
      \multicolumn{1}{c}{} &
       &
      \multicolumn{1}{c}{\cellcolor[HTML]{FFFFFF}} &
      \multicolumn{1}{c}{} \\
    \multirow{-3}{*}{} &
      \multicolumn{1}{c}{\multirow{-2}{*}{\cellcolor[HTML]{FFFFFF}Manual}} &
      \multicolumn{1}{c}{Initial} &
      \multicolumn{1}{c}{Refined} &
      \multicolumn{1}{c}{\multirow{-2}{*}{\cellcolor[HTML]{FFFFFF}Manual}} &
      \multicolumn{1}{c}{Initial} &
      \multicolumn{1}{c}{Refined} &
       &
      \multicolumn{1}{c}{\multirow{-2}{*}{\cellcolor[HTML]{FFFFFF}Manual}} &
      \multicolumn{1}{c}{\multirow{-2}{*}{Code}} &
       &
      \multicolumn{1}{c}{\multirow{-2}{*}{\cellcolor[HTML]{FFFFFF}Manual}} &
      \multicolumn{1}{c}{\multirow{-2}{*}{Code}} \\ \midrule
     &
       &
      \multicolumn{1}{l}{} &
      \multicolumn{1}{l}{} &
       &
      \multicolumn{1}{l}{} &
      \multicolumn{1}{l}{} &
       &
       &
      \multicolumn{1}{l}{} &
       &
       &
      \multicolumn{1}{l}{} \\
    \multicolumn{13}{l}{\textbf{\underline{Annotation 1:} Webpack Bundle}} \\
    \# Webpack Bundles &
      \multicolumn{1}{r}{104} &
      153 &
      111 &
      \multicolumn{1}{r}{156} &
      162 &
      163 &
       &
      \multicolumn{1}{r}{207} &
      217 &
       &
      \multicolumn{1}{r}{99} &
      102 \\
    % Error Rate &
    %    &
    %   0.32 &
    %   0.06 &
    %    &
    %   0.06 &
    %   0.04 &
    %    &
    %    &
    %   0.05 &
    %    &
    %    &
    %   0.03 \\
    Precision &
       &
      0.68 &
      0.94 &
       &
      0.94 &
      0.96 &
       &
       &
      0.95 &
       &
       &
      0.97 \\
    \cellcolor[HTML]{FFFFFF}Recall &
       &
      1 &
      1 &
       &
      1 &
      1 &
       &
       &
      1 &
       &
       &
      1 \\
    \cellcolor[HTML]{FFFFFF} &
       &
      \multicolumn{1}{l}{} &
      \multicolumn{1}{l}{} &
       &
      \multicolumn{1}{l}{} &
      \multicolumn{1}{l}{} &
       &
       &
      \multicolumn{1}{l}{} &
       &
       &
      \multicolumn{1}{l}{} \\
    \multicolumn{13}{l}{\textbf{\underline{Annotation 2:} Component Modules}} \\
    \# Modules &
      \multicolumn{1}{r}{6,857} &
      7,847 &
      6,857 &
      \multicolumn{1}{r}{9,816} &
      9,156 &
      9,090 &
       &
      \multicolumn{1}{r}{12,889} &
      12,913 &
       &
      \multicolumn{1}{r}{5,701} &
      5,690 \\
    % Error Rate &
    %    &
    %   0.14 &
    %   0.002 &
    %    &
    %   0.007 &
    %   0.002 &
    %    &
    %    &
    %   0.002 &
    %    &
    %    &
    %   0.002 \\
    Precision &
       &
      0.87 &
      0.99 &
       &
      0.99 &
      0.99 &
       &
       &
      0.998 &
       &
       &
      0.998 \\
    Recall &
       &
      1 &
      1 &
       &
      0.93 &
      1 &
       &
       &
      1 &
       &
       &
      0.996 \\ \bottomrule
    \end{tabular}%
    }
    \end{table*}

\subsection{Gathering Modules}
\label{sec:gathering-modules}

\system first generates an Abstract Syntax Tree (AST) representation of a \js resource. The AST representation helps gather the syntactic features of the script and provides insight into the structure of the code. 

Thereafter, \system uses the AST representation to determine two aspects of the loaded script. First, \system determines if the script comprises a bundle with one or more modules. Second, \system gathers the sub-trees corresponding to each identified module.

Below, we describe the webpack-specific implementation.

\subsubsection{Gathering \js resources}
To create a valid parser for bundles in the wild, we gathered examples of resources loaded on popular websites. We developed a puppeteer-based crawler that, upon visiting a page, intercepts network requests and stores a copy of observed script responses. We visited domains from the Tranco list~\cite{TrancoPaper} and gathered 30,930 scripts from 1,063 sites (1K crawl).

\subsubsection{Generating ASTs}
For each of the gathered scripts, \system uses acorn, a community-developed, open-source \js parser to generate an AST\footnote{\url{https://github.com/acornjs/acorn}}. The generated AST complies with the ESTree Spec to ensure a consistent, standardized representation that can be reproduced by alternative implementations\footnote{\url{https://github.com/estree/estree}}.

\subsubsection{Code Development and Refinement Methodology.}
We began with an understanding of the general structure of the output of a webpack bundle (see Listing~\ref{lst:example-bundled-code}). We developed a script that parses an AST and looks for a module array or object, i.e., an array or object that comprises functions. We used this initial logic and adopted a code optimization methodology based on ground truth gathered through expert manual annotation. We describe the process below.

\begin{enumerate}[leftmargin=*,nosep,noitemsep]
    \item First, we randomly sampled 100 scripts (without replacement) from the \js resources gathered during the 1K site crawl. 
    \item We manually evaluated the plaintext script and the AST of each sampled resource. In doing so, we checked for the presence of webpack-specific code patterns. We annotated each resource with a boolean value indicating whether we determined the script as a webpack bundle (\textbf{Annotation 1}). Note that these manual checks (\eg keywords) were only used for annotation and not incorporated into the automated approach. The manual checks included: 
    \begin{enumerate}
        \item identification of code pattern similar to webpack's specific function that handles dependencies (see Listing~\ref{lst:example-bundled-code}, L6-26);
        \item identification of code patterns specific to webpack chunks, i.e., files comprising webpack modules separate from the entry point bundle;
        \item identification of objects and arrays specific containing functions with parameters, exports, and return statements similar to webpack's function wrappers;
        \item keyword searches that indicate the use of webpack in the use of webpack in the creation of the resource.
    \end{enumerate}
    \item For any resource annotated as a webpack bundle, we also annotated the resource with the number of component modules we identified (\textbf{Annotation 2}).
    \item We repeated Steps 1-3 until at least we had annotated 100 scripts as webpack bundles.
    \item We then executed our code to automatically analyze and annotate each sampled resource in a similar manner, \ie{} (a) \textbf{Annotation 1}: whether the script is a bundle, and (b) \textbf{Annotation 2}: the number of component modules identified in the script. 
    \item We compared the code-generated annotations against the manual annotations and gathered the set of differences. For each incorrectly labeled script, we analyzed the code outputs and refined its logic. Refinements included addressing edge cases, handling multiple IIFEs in a single resource, etc.
    \item We used the refined version of our code to annotate the same sample and noted any improvements. This refined code version is the initial version of the subsequent phase against new samples.
    \item For each subsequent round, we repeated Steps 1-7, appending the existing sample with 50 manually annotated webpack resources (Step 4). We stopped refining our logic when we observed negligible improvement in the precision, recall, and accuracy of code-generated annotations between consecutive phases.
\end{enumerate}

We repeated the process for three rounds. Table~\ref{tab:bundle-sampling} presents the numbers and metrics for both annotations and the associated metrics for each round of refinement. At the end of the third round, our code identified a given script as a webpack bundle with 95\% precision and gathered modules with 99.8\% precision. 

We used the resulting logic to annotate all scripts gathered in our 1K crawl. We annotated a total of 11,995 scripts as webpack bundles. Finally, we performed a spot check of the generated annotation. We randomly sampled and manually annotated 300 previously unsampled scripts, of which we identified 99 scripts as webpack bundles. Upon comparing and verifying the code-assigned annotation, we observed a 97\% precision rate in identifying webpack bundles and a 99.8\% precision rate in gathering component modules.

\begin{algorithm}[t]
    \caption{Processing modules.}\label{alg:preprocessing-modules}
    \begin{algorithmic}
        \State \underline{\textbf{Input:}} AST $\gets$ module AST
        \State \underline{\textbf{Output:}} hashedAST $\gets$ processed representation of the AST
        \Procedure{hashModule}{$node$}
            \State $childrenHash\gets 0$
            \ForEach{$child$ in $node.children$}
            \State $childrenHash\gets childrenHash +\Call{hashModule}{child}$ 
            \EndFor
            \State \textbf{return} $hash(node.type + childrenHash)$ 
        \EndProcedure
    \end{algorithmic}
\end{algorithm}

\subsection{Processing Modules}
\label{sec:processing-modules}

Each module gathered from the previous phase comprises a sub-tree, i.e., a partial AST of the larger AST representation of the script. The module's AST contains information about the structure of the underlying code and the associated names of variables, functions, and properties included within the script. Script attributes like variable names are extremely volatile and have limited use in reliably identifying target modules. \system, therefore, only considers (1) the \textit{structure} of the AST, i.e., the parent-child relationships between the nodes that comprise the AST, and (2) the \textit{type} of each AST node, i.e., attributes which comply with the ESTree Spec, and are hence limited to a deterministic set of values.

To distill and represent only these specific attributes from a module's AST, we adopt a version of a cryptographic concept used in integrity verification: Merkle trees~\cite{Merkle:1987}. This approach presents a novel part of our methodology. For example, consider a list of data blocks that need to be verified (e.g., transactions in a blockchain). We represent this list of blocks in a tree structure. Consider a tree where each leaf node represents a piece of data, and each non-leaf node is a cryptographic hash of its child nodes. These hashes propagate upwards, converging into a singular root hash known as the Merkle root. Any alteration in the foundational data triggers a modification in the root hash, instantly signaling tampering or modifications. However, if the root hash is verified, all of its children are also verified.

\system adopts a similar methodology, briefly presented in Algorithm~\ref{alg:preprocessing-modules}. Given a module's AST, it performs the following processing steps.
\begin{enumerate}[leftmargin=*,nosep,noitemsep]
    \item \system begins traversing the module AST from the root node.
    \item For every \textit{child node}, it recursively calls the function to gather the \textit{child node}'s hash. The hash of each child, in turn, results from the hashes of its children.
    \item It sums the hashes of all \textit{child nodes} of the root node.
    \item To generate a hashed representation of the root node, \system concatenates the \textit{type} of the node to the sum of hashes of its children and hashes the resulting string.
    \item It returns the processed representation of the AST.
\end{enumerate}

This processed representation now comprises an alternative tree representation with equal depth to its original counterpart. However, tree-based comparisons are complex and add a large performance overhead.
Recall that while processing an AST, top-level nodes contain information about underlying children. Therefore, comparing a limiting comparison to a few high-level nodes of the processed module can provide insight into the similarity of the module's AST. To this end, \system gathers a list of top-level nodes. In other words, noting that webpack wraps each module within a function, we consider the computed hashes of each statement within the function. Say, for example, when bundled within webpack, the module comprises three statements: one variable declaration statement, one function definition (which further contains multiple nested statements), and one export statement. \system will process this module and gather hashes for three top-level nodes -- owing to the Merkle tree-based approach, each of the hashes will be computed from the top-level statements' children. 

Additionally, \system associates each entry in this list with a weight that represents the number of child nodes it represents, i.e., it weighs a node representing a complex function higher than a node representing a simple variable declaration. Consider the same example as above. Here, the initial declaration statement and the export statement will represent fewer nodes than the function definition (which has children within nested statements). As a result, the function declaration statement will have a higher weight, corresponding to the ratio of nodes it represents to the total nodes in the module's overall AST. 

\begin{algorithm}[t]
\caption{Comparing modules against target libraries.}\label{alg:compare-modules}
\begin{algorithmic}
    \State \underline{\textbf{Input:}}
    \State TOP\_LEVEL\_NODES $\gets$ list of top-level nodes of module 
    \State DATABASE $\gets$ hashes of known libraries 
    \State CANDIDATE\_LIBRARIES $\gets$ empty set
    \State \underline{\textbf{Output:}}
    \State MATCH $\gets$ weight of the associated match
    \ForEach{$nodeHash$ in $TOP\_LEVEL\_NODES$}
        \State librariesWithThisNode = DATABASE[$nodeHash$]
        \ForEach{[library, weight] in librariesWithThisNode}
            \State CANDIDATE\_LIBRARIES[library] $+$$=$ weight 
        \EndFor
    \EndFor
    \State bestMatch = max(CANDIDATE\_LIBRARIES)
    \If{bestMatch = targetLibary}
        \Return CANDIDATE\_LIBRARIES[targetLibrary]
    \EndIf
\end{algorithmic}
\end{algorithm}

\subsection{Comparing Modules}
\label{sec:comparing-modules}

\system uses the list of top-level nodes to compare against a database of top-level nodes of processed module representations corresponding to targeted libraries. While this database is generated offline, processed modules for these libraries are also gathered in a similar manner to the logic presented in Algorithm~\ref{alg:preprocessing-modules}. We provide details regarding the creation of this database in Section~\ref{sec:evaluation}. Here, we focus our discussion on the comparison and identification of processed modules.

Algorithm~\ref{alg:compare-modules} presents an overview of the comparison strategy. We present a brief description below.
\begin{enumerate}[leftmargin=*,nosep,noitemsep]
    \item \system traverses the list of hashes of the top-level nodes representing the module observed in the wild.
    \item It looks up the hash of each node in the database and identifies every library with the same top-level node. It considers any such library as a \textit{candidate library}.
    \item It appends the node's weight for each \textit{candidate library} to previous matches, if any. Therefore, the weights associated with a \textit{candidate library} increase each time it includes a match.
    \item Once all top-level nodes have been traversed, \system identifies the \textit{candidate library} with the highest associated weight.
    \item If the highest match corresponds to the target library, \system returns the weight of this match, thereby marking the module as a candidate for replacement. 
\end{enumerate}

\subsection{Replacing Modules}

\system replaces each marked privacy-harming module with a corresponding, benign replacement. It can perform this action in one of two ways: (1) \system can replace the module's AST with an alternative, benign AST and then regenerate the script from the modified AST; (2) Alternatively, \system can identify the string indices for the module within the textual representation of the script, and place the benign replacement between these indices. Regardless of its approach, \system ensures that access and use of the replacement does not break other parts of website functionality.

Replacements for targeted modules are manually created. Given the nuances of each target library and its eventual mutated representation in webpack bundles, we describe the three categories of replacements to consider within each module.

\begin{itemize}[leftmargin=*,nosep,noitemsep]
    \item[\ding{228}] \textbf{Global Variables.} If libraries expose global variables on the \texttt{window} object, these variables must be made available. Additionally, the types of such variables need to be retained.
    \item[\ding{228}] \textbf{Exports.} The replacement must ensure that values previously exported from the target module remain available. When other modules consume the module, it must be made accessible, even if replaced with a benign version.  
    \item[\ding{228}] \textbf{Webpack-based Function Wrappers.} When webpack wraps modules in functions, it passes specific parameters included in the function signature. Additionally, values exported by modules are appended as properties to webpack objects. Replacements need to ensure that these parameters and exports work seamlessly. 
\end{itemize}

Finally, each replacement needs to ensure the consistency of variable types, i.e., functions be replaced with functions, constants be replaced with constants, and so on. We further provide examples of specific replacements in Section~\ref{sec:evaluation}.

%% file: 04-evaluation.tex
\section{Evaluation}
\label{sec:evaluation}
%-------------------------------------------------------------------------------

In this section, we evaluate the effectiveness of \system in identifying specific libraries in bundled scripts captured in the wild. We select three libraries to target and describe the process of gathering their module representations for comparison. We then use \system to identify target libraries in scripts gathered from a crawl of popular websites. Next, we develop a deployment of \system as a Firefox extension and evaluate the performance overhead introduced by its interventions. We conclude by describing the creation and evaluation of benign replacements.

\subsection{Evaluation Dataset}
\label{sec:evaluation-dataset}

We first describe the selection of example libraries that we use to identify in the wild. We describe the offline process of gathering representations for these libraries, which \system can then use to identify libraries in the wild.

\subsubsection{Target Library Selection}
We select three libraries for our evaluation \textemdash FingerprintJS\footnote{\url{https://github.com/fingerprintjs/fingerprintjs}}, Sentry\footnote{\url{https://github.com/getsentry/sentry-javascript}}, and Prebid\footnote{\url{https://github.com/prebid/Prebid.js}}. Each of these libraries is included in filter lists\footnote{\url{https://easylist.to/}} used by popular content blocking tools\footnote{https://github.com/gorhill/uBlock}\footnote{\url{https://adblockplus.org/}}. Additionally, all three libraries provide the option for developers to use their \texttt{npm} packages with bundled code.

\textbf{FingerprintJS}\footnote{\url{https://fingerprint.com/}} is a popular browser fingerprinting library that tracks and identifies users. When embedded within a website that a user visits, the library performs various client-side operations that query browser attributes and stores a unique identifier for the user. Besides its use in user authentication~\cite{Lin:2022} and ad fraud prevention\footnote{\url{https://fingerprint.com/case-studies/}}, the library de-anonymizes sensitive user activity across sites. Since the domains that the library uses are blocked by popular content-blocking tools, FingerprintJS recommends that developers use their \texttt{npm} package or self-host a copy of their scripts\footnote{\url{https://github.com/fingerprintjs/fingerprintjs/blob/master/docs/evade_ad_blockers.md}}.

\textbf{Sentry}\footnote{\url{https://sentry.io/}} is an analytics tool that offers libraries for performance and error monitoring. The library helps developers gather details about user interaction, DOM events, console logs, and network calls. The library lets developers decide how they use the information gathered from users and offers multiple integrations, including multiple third-party analytics services\footnote{\url{https://develop.sentry.dev/development/analytics/}}. Since the CDNs that host Sentry code are blocked by popular content blockers, the library recommends that developers get around this restriction by bundling Sentry's \texttt{npm} package into their app\footnote{\url{https://docs.sentry.io/platforms/javascript/troubleshooting/\#dealing-with-ad-blockers}}.

\textbf{Prebid}\footnote{\url{https://prebid.org/}} is a popular advertising library that developers can use to add header bidding to their applications. The library integrates with numerous advertising, analytics, and user-tracking libraries, providing support for bidding on targeted advertisements. Similar to previously discussed libraries, Prebid is included within filter lists and content-blocking tools. The library is open source and is available to be added to bundles with an \texttt{npm} package. Prebid provides specific instructions on integration with webpack\footnote{\url{https://github.com/prebid/Prebid.js}}. The library must be compiled with Babel\footnote{\url{https://babeljs.io/}}, and specific plugins must be used when loaded with webpack. It provides specific instructions for webpack configurations, which gives us insight into its use when integrated by bundles in the wild.

% Please add the following required packages to your document preamble:
% \usepackage{booktabs}
% \usepackage{graphicx}
\begin{table}[t]
    \centering
    \caption{An overview of the build options used to gather representations for the target library.}
    \label{tab:configuration-options}
    % \resizebox{\columnwidth}{!}{%
    \begin{tabular}{@{}lc@{}}
    \toprule
    \textbf{Build Options}                   & \textbf{Possible Values}   \\ \midrule
    \textbf{Module Systems:} &  \\
    $\hookrightarrow$ Dependency Statement            & {[}import, require{]} \\
    \textbf{Webpack Optimizations:} &                   \\
    $\hookrightarrow$ usedExports                     & {[}true, false{]} \\
    $\hookrightarrow$ concatenatedModules             & {[}true, false{]} \\
    \textbf{Minifier Options:} &  \\
    $\hookrightarrow$ passes                          & {[}0, 1, 2{]}  \\
    $\hookrightarrow$ arrows                          & {[}true, false{]} \\
    $\hookrightarrow$ dropConsole                     & {[}true, false{]} \\
    $\hookrightarrow$ unsafeCompress                  & {[}true, false{]} \\
    $\hookrightarrow$ unsafeMethods                   & {[}true, false{]} \\ 
    $\hookrightarrow$ unsafeUndefined                 & {[}true, false{]} \\
    $\hookrightarrow$ unsafeArrow                     & {[}true, false{]} \\
    $\hookrightarrow$ unsafeCompare                   & {[}true, false{]} \\
    $\hookrightarrow$ typeofs                         & {[}true, false{]} \\
    \textbf{Browser Support:}  & \\ 
    $\hookrightarrow$ ie8                             & {[}true, false{]} \\
    $\hookrightarrow$ safari10                        & {[}true, false{]} \\
    \bottomrule
    \end{tabular}%
    % }
\end{table}

\subsubsection{Bundler Configurations.} Web applications are developed in a large variety of environments, frameworks, and configurations before they are compiled and shipped to production. As a result, when the same \texttt{npm} package is bundled within different applications, its modules will generate a wide range of AST structures. These variations make it difficult to gather a definitive AST representation for the package that will return perfect matches when compared against scripts found in the wild.
However, the AST of the module corresponding to each library is complex and, hence, unique. Our intuition is that the AST of a given library will be closer to differently configured ASTs of the same library than to ASTs of other libraries. 

To gather multiple options for module representations and help us get the closest match to a target library in the wild, we gather a list of popular build options to be considered for each library. Table~\ref{tab:configuration-options} These options cover four aspects that we describe below.

\begin{enumerate}[leftmargin=*,nosep,noitemsep]
    \item \textbf{Module Systems.} Depending on the target \texttt{npm} package and the development environment of the web application, the library may be consumed as a CommonJS module or an ECMAScript module. We gather configurations for both module types.
    \item \textbf{Webpack Optimizations.} When bundling code from multiple files, webpack keeps track of imported and exported values with the help of a dependency graph. Depending on the module system, the output environment, and the developer's configurations, it provides options to automatically identify and discard unused parts of code and to reduce the size of the final output. The use of these options can change how the module appears in the output bundle.
    \item \textbf{Minifier Options.} Once webpack creates a bundle output, it uses Terser\footnote{\url{https://terser.org/}} to minify the code. Developers can provide additional options to ensure that the output works with their intended production environment and that their applications work in all supported environments. We consider a list of compression options that alter the syntax of the module in the bundled output.
    \item \textbf{Browser Support.} Developers can additionally specify support for legacy browsers. These options override other minification and compression to ensure that the output bundle is also compatible with legacy browsers.
\end{enumerate}

We gathered all combinations of the build options and created a total of 24,576 build configurations that we then used to bundle each target library.

\subsubsection{Creation of Target Module Representation Database.}

We gather various hashes for these libraries with the help of a large set of versions and different webpack configuration options, resulting in multiple AST representations for each target library. We briefly describe the steps below.

\begin{enumerate}[leftmargin=*,nosep,noitemsep]
    \item \textbf{Library Versions.} We gather a list of past versions released by the library on the \texttt{npm} registry. We download and install each version in a separate barebones application. 
    \item \textbf{Barebones Application.} For each version, we create a barebones Node.js application with a single \js file. The file consumes the target library with either an \texttt{import} or a \texttt{require} statement.
    \item \textbf{Build Options.} For each library version, we build multiple webpack bundles, one for each combination of build options.
    \item \textbf{Target Module.} When building a bundle, webpack provides the option to gather relevant info about individual modules\footnote{\url{https://webpack.js.org/api/stats/}}. We use this information to identify the file, name, and position of the module corresponding to the target library. Thereafter, we gather an AST representation of the output bundle and extract the sub-tree corresponding to the target module.
    \item \textbf{Processing AST.} We process the extracted AST in a similar manner to the steps described in Section~\ref{sec:processing-modules}. We then gather a list of hashes for the top-level nodes from each bundle along with their corresponding weights. 
\end{enumerate}

\subsubsection{Additional Library Configurations.}
\label{sec:additional-configurations}
In addition to the three target libraries, we gathered a list of 10k popular \texttt{npm} packages based on their download counts\footnote{\url{https://github.com/anvaka/npmrank}}. From this list, we collected a random sample of 10 versions of each of the 1k libraries. For each sampled library version, we generated 100 webpack bundles each with a different, random build option. We extracted and processed their modules and gathered a weighted list of hashes corresponding to the top-level nodes of their ASTs.

\subsection{Framework Effectiveness}

In this section, we describe our evaluation of \system's effectiveness in identifying target libraries within scripts captured in the wild.

\subsubsection{\js Resource Dataset.} We used a puppeteer-based crawler that visited sites from the Tranco list~\cite{TrancoPaper} and gathered \js resources through network interception. We previously used this dataset to create a bundle parser, described in Section~\ref{sec:gathering-modules}. The dataset comprises 30,930 scripts from 1,063 sites, of which 11,995 scripts are bundled resources.

\subsubsection{Gathering Matches.} For each bundle in the dataset, \system extracted and processed component modules (see Algorithm~\ref{alg:preprocessing-modules}). It then gathered a weighted list of top-level nodes from the processed module. \system then compared each module against all libraries in the evaluation dataset (see Section~\ref{sec:evaluation-dataset}). For each module, we noted the library in the evaluation dataset with the closest match. Finally, we gathered a list of modules for which the closest match was one of the three target libraries. We gathered a total of 73 matches for FingerprintJS, 324 matches for Sentry, and 3,532 matches for Prebid.

\subsubsection{Manual Verification.} For each match, we manually verified the code embedded between the corresponding indices in the script. We observed the code structure, the use of specific properties, and the number and types of exported variables, functions, and objects. We then manually annotated each match as a true positive or a false positive.

% \begin{table}[t]
%     \centering
% \caption{Matches observed for each target library in the evaluation dataset ($n=1,063$ sites).}
% \label{tab:target-library-detection}
%     % \resizebox{\columnwidth}{!}{%
%     \begin{tabular}{lrr}
%         \toprule
%          \textbf{Target Library}&  \textbf{\# Matches}& \textbf{\# Matches > 8\%}\\
%          \midrule
%          FingerprintJS &  73  &  40\\
%          Sentry &  324  &  54 \\
%          Prebid & 3,532  &  46 \\
%         \bottomrule 
%     \end{tabular}
%     % }
% \end{table}

\subsubsection{True Positive Threshold.} We observed the tradeoff between precision and recall for the matches returned for all three target libraries. We looked for the lowest threshold match percentage for an AST for which all three libraries observed 100\% precision, \ie{} any AST that returned a positive match above this threshold was correctly annotated. We, therefore, arrived at a common match threshold value of 8\%. Despite observing a small drop in recall for Sentry, we prioritized precision to ensure that \system never blocks benign modules even if it misses some privacy-harming scripts. 
% Across all three libraries, we noted that \system correctly identified modules that shared >=8\% of their nodes with the target module. Table~\ref{tab:target-library-detection} provides a snapshot of the matches. \system correctly identified 40 instances of FingerprintJS, 54 instances of Sentry, and 46 instances of Prebid in the dataset. \kaytwo{might be worth mentioning here that this was done with no false positives (if that's true, which is my recollection)}

\begin{figure}[t]
    \centering
    \includegraphics[width=0.98\columnwidth,keepaspectratio]{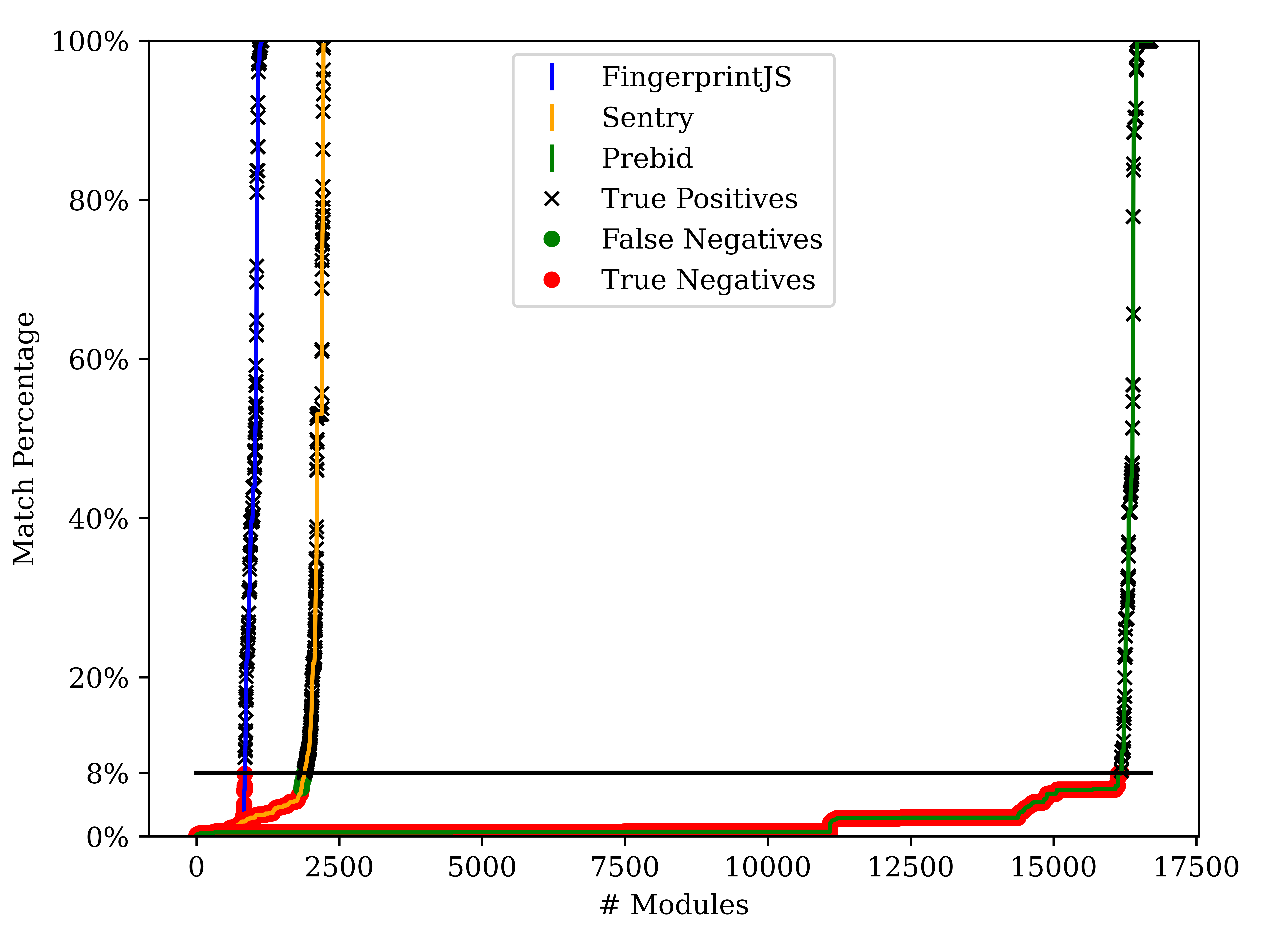}
    \vspace{-3ex}
    \caption{An overview of the target libraries identified from a crawl of the Tranco 10K. In addition to the 697 instances of true positive (\ding{53}) matches above the 8\% threshold, we observed $>$15K instances of true negative (\tikzcircle[red, fill=red]{1.5pt}) matches of modules below the threshold.}
    \label{fig:match-cdf}
    \vspace{-2ex}
\end{figure}

\subsubsection{Evaluation on a larger crawl.} 
\label{sec:10k-crawl}
Next, we performed a larger crawl of the web, gathering scripts from the top 10K sites in the Tranco list. We used \system to evaluate these scripts and employed the >=8\% match threshold discussed earlier. We detected FingerprintJS on 205 sites, Sentry on 213 sites, and Prebid on 325 sites. Overall, \system identified at least one of the three target libraries on 7\% ($n=697$) of the top 10K sites. Figure~\ref{fig:match-cdf} presents an overview of our observations. We note that the larger crawl has fewer matches than the Top-1K sites, which can possibly be explained by existing work, which states that (a) less popular websites are less complex~\cite{ruth2022}, and (b) popular sites are more likely to engage in privacy-harming behaviors~\cite{Papadogiannakis2021}.

\subsection{Performance}

In this section, we provide a sample deployment of \system as a Firefox extension. We use this non-optimized deployment to evaluate an upper bound on the latency introduced by different phases of the pipeline and their effect on page load times.

\subsubsection{Firefox Extension.} 
Considering \system's ability to recognize bundles given the entire contents of scripts, we determined that an approach that intercepts network requests and responses will be a useful initial deployment. Considering Chrome's restrictions on content filtering from extensions\footnote{\url{https://developer.chrome.com/blog/improvements-to-content-filtering-in-manifest-v3/}}, we developed a Firefox extension that intercepts and modifies the content of responses across all domains that the browser visits\footnote{\url{https://blog.mozilla.org/addons/2022/05/18/manifest-v3-in-firefox-recap-next-steps/}}.

% Upon first loading the browser, the extension gathers a precompiled dataset of known library hashes (see Section~\ref{sec:evaluation-dataset}) and creates a hashmap for all future comparisons.

\subsubsection{Crawl.}
\label{sec:performance-crawl}
We evaluated the performance of a browser instance with and without the extension by visiting sites in the Tranco list~\cite{TrancoPaper}. For each web page, we performed the following steps.
\begin{enumerate}[leftmargin=*,nosep,noitemsep]
    \item First, we created and initialized a new browsing profile.
    \item Next, we used Mozilla's \texttt{web-ext}\footnote{\url{https://extensionworkshop.com/documentation/develop/getting-started-with-web-ext/}} to load a Firefox instance with the web extension.
    \item We used Puppeteer\footnote{\url{https://pptr.dev/}} to connect to the browser instance, visit the web page in a new tab and wait for 30 seconds.
    \item The extension's background script intercepts and evaluates all script-based network requests. It captures and stores the time taken by each step in the script evaluation pipeline. 
    \item The extension's content script captures metrics relevant to web page performance.
    \item We closed the browser instance and deleted the user profile directory and Firefox's caches from the filesystem to ensure a fresh browsing state for subsequent visits.
    \item We repeated steps 1-6 in a browsing profile without our extension installed and recorded relevant web page performance metrics for comparison.
\end{enumerate}

\begin{table}[t]
    \centering
\caption{Time taken (ms) by the extension to process scripts using network interception.}
\label{tab:performance-phases}
    % \resizebox{\columnwidth}{!}{%
    \begin{tabular}{lrrrr}
        \toprule
        \multirow{2}{*}{\textbf{Phase}}&  \multirow{2}{*}{\textbf{\# Scripts}} & \multicolumn{3}{c}{\textbf{Time (ms)}}\\
         & &  \multicolumn{1}{c}{\textbf{$\mu$}} &
         \multicolumn{1}{c}{\textbf{$\sigma$}} &
         \multicolumn{1}{c}{\textbf{Median}} \\ 
         \midrule
         Buffering script text &  6,702 & 690.69 & 1,189.26 & 240 \\
         \midrule
         Gathering modules & 6,699 & 46.17 & 98.25 & 9 \\
         Processing modules & 2,135 & 326.31 & 635.58 & 114 \\
         Comparing modules & 2,135 & 1.15 & 2.28 & 0 \\
         \bottomrule
    \end{tabular}
    % }
\end{table}

\subsubsection{Pipeline Performance.}
\label{sec:performance-pipeline}
First, we discuss the evaluation of individual scripts. Table~\ref{tab:performance-phases} shows the time taken by different phases of the pipeline. We observed the gathering modules and compared processed modules, which have the least impact, taking, on average, 46.17ms and 1.15ms, respectively. \system takes the longest time to process modules, i.e., recursively compute hashes for all nodes in an AST, which takes 326.31ms on average for each script. 

However, we observed that the largest chunk of the time was consumed outside of \system's evaluation and instead spent in buffering incoming response chunks to load the entire script. This is a limitation of the mode of deployment, i.e., Firefox extensions that modify the body of network responses bypass the browser's optimized cache for scripts\footnote{\url{https://developer.mozilla.org/en-US/docs/Mozilla/Add-ons/WebExtensions/API/webRequest/filterResponseData}}. The deployment waits for large network responses to complete before evaluating the script. In Section~\ref{sec:discussion}, we discuss alternative deployments that evaluate code at runtime and can bypass dependence on network load times.

% Please add the following required packages to your document preamble:
% \usepackage{booktabs}
% \usepackage{graphicx}
% \usepackage[normalem]{ulem}
% \useunder{\uline}{\ul}{}
\begin{table}[t]
\centering
\caption{Time taken (ms) to load a page with and without the extension installed, from a crawl of $n=963$ web pages.}
\label{tab:performance-pageload}
\resizebox{\columnwidth}{!}{%
\begin{tabular}{@{}lrrrrrr@{}}
\toprule
\multirow{2}{*}{\textbf{Metric}} & \multicolumn{3}{c}{\textbf{w/o Extension}} & \multicolumn{3}{c}{\textbf{w/ Extension}} \\ 
\multicolumn{1}{c}{\textbf{}} &
  \multicolumn{1}{c}{\textbf{$\mu$}} &
  \multicolumn{1}{c}{\textbf{$\sigma$}} &
  \multicolumn{1}{c}{\textbf{Median}} &
  \multicolumn{1}{c}{\textbf{$\mu$}} &
  \multicolumn{1}{c}{\textbf{$\sigma$}} &
  \multicolumn{1}{c}{\textbf{Median}} \\ \midrule
First Contentful Paint              & 2,735.55      & 1,690.62      & 2,550      & 3,029.51      & 2,236.36      & 2,423     \\
DOM Interactive                     & 2,796.39      & 1,687.63      & 2,620      & 3,367.55      & 2,204.51      & 3,052     \\
Page Load                           & 4,790.39      & 3,156.72      & 4,130      & 7,107.45      & 4,759.42      & 6,078     \\ \bottomrule
\end{tabular}%
}
\end{table}

\begin{table}[t]
\centering
\caption{Memory space taken by the extension's hash signatures to detect target libraries, comprising signatures of target libraries along with those of additional libraries (\S\ref{sec:additional-configurations}).}
\label{tab:database-size}
\resizebox{\columnwidth}{!}{%
\begin{tabular}{@{}llrrr@{}}
\toprule
\multicolumn{2}{c}{\textbf{Signatures}} &
  \multirow{2}{*}{\textbf{\#ASTs}} &
  \multicolumn{1}{c}{\textbf{\#HashMap}} & 
  \multirow{2}{*}{\textbf{Size}} \\
\multicolumn{1}{c}{\textbf{Target}} &
  \multicolumn{1}{c}{\textbf{Addtional}} &
  \multicolumn{1}{c}{} &
  \multicolumn{1}{c}{\textbf{Entries}} & 
  \multicolumn{1}{c}{} \\ \midrule
All Hashes              & All Hashes      & 258,788      & 265,962      & 3.55 GB     \\
All Hashes              & 0.5 * (All Hashes)      & 176,963      & 142,173      & 2.77 GB     \\
All Hashes              & \textemdash            & 95,174      & 19,555      &   2.27 GB    \\ 
Top-10K Matches      & \textemdash            & 242          & 2,514       & 0.003 GB \\  
\bottomrule
\end{tabular}%
}
\end{table}

\subsubsection{Page Performance.}
Next, we discuss the effect of the sample deployment on page load. Recall that we used a content script to gather metrics from the page context for each visit with and without the extension. Table~\ref{tab:performance-pageload} presents a snapshot of our observations. We describe three collected measures below.

\textit{First Contentful Paint} (FCP) is a timing measure that shows when the browser renders the first bit of content. The content could be any text, image, video, canvas, or non-empty SVG. The timing shows the first instance in which the user has an indication that the page is loading. We observed that the extension added 293.36ms to the average time a user would wait for such an indication.

\textit{DOM Interactive} indicates the time taken for the Document Object Model (DOM) parser to finish its work on the main document, i.e., the time taken to construct the DOM. This time can be affected by the parser-blocking JavaScript. Note that the extension's script evaluation runs outside the main thread. We observed that the extension added 571.66ms on average, i.e., a user would have to wait an additional 0.5s before the browser has parsed the DOM.

\textit{Page Load} indicates the total time taken for all resources to load. It indicates that network requests for all scripts, images, and other resources have been completed. This metric does not indicate actual end-user experience since it depends on device capabilities and various network conditions, but in our case, provides insight into the additional time added by the extension buffering responses and evaluating script contents. We observed that pages loaded with the extension took an additional 2,317.06ms on average before triggering the \texttt{load} event.

\subsubsection{Signature Database Size.}
\label{sec:database-size}
\system's extension deployment uses a complete database of hash signatures, which includes those computed for target library configurations (based on Table~\ref{tab:configuration-options}) along with additional configurations of popular NPM libraries (see \S\ref{sec:additional-configurations}). To optimize the time taken to compare modules (\S\ref{sec:comparing-modules}), URR stores the signatures as a hashmap, with the key for each entry holding the hash of a \textit{top-level} node and the value holding a dictionary comprising \textit{root node} hashes along with corresponding weights (see Algorithm~\ref{alg:compare-modules}). We evaluated different database compositions to reduce its overall size, especially considering subsets of additional configs from popular libraries. Table~\ref{tab:database-size} shows that even with the database solely comprising signatures of target libraries, it consumes 2.27GB of memory space (see Row 3 of Table~\ref{tab:database-size}). We observed that reducing the number of hashes included in the database reduced the extension's use of memory space (from 3.55 GB to 2.27 GB), which can be beneficial to devices with limited resources.  

We considered further reducing the database to only include target library signatures we found in our 10K crawl (see \S\ref{sec:10k-crawl}). While this subset greatly reduced the size of the database (3.54 MB), we noted that the majority of the  ASTs (57\%; $n=139$) correspond to a single match. We comprised an extensive list of hash representations (based on Table~\ref{tab:configuration-options}) to make URR especially effective in detecting target libraries despite the websites adopting of a wide array of potential configurations. Using a match-based subset (as used in Row 4 of Table~\ref{tab:database-size}) will limit URR's detection of libraries on websites that deploy target libraries within differently configured webpack bundles. We note that while greatly reducing target library hash representations may not be effective, developers can consider caching frequently encountered bundles. We further discuss caching in \S\ref{sec:caching}.

Finally, we repeated crawls of sites from the Tranco list (see \S\ref{sec:performance-crawl}), but this time, visited sites with each of the four database sizes deployed within URR. We observed that the different configurations had a negligible effect on page load, with the databases' use in the \textit{Comparing Modules} phase taking an average time of 1ms ($M=0$ms). Table~\ref{tab:database-size} therefore focuses on the different configurations' impact on memory usage. 

% Overall, deploying \system as a browser extension slightly increases the time taken to load the page, adding minimal performance overhead and ensuring that users can continue to meaningfully interact with the website.

\subsection{Replacements}
\label{sec:evaluation-replacements}

Prior work has extensively studied and determined the creation and use of non-breaking, benign replacements for privacy-harming code~\cite{SugarCoat:2021}. We follow similar principles and extend example replacements used by uBlock Origin\footnote{\url{https://github.com/gorhill/uBlock/tree/9123563895f0499849b4d85c4f95e1ed6ace2231/src/web_accessible_resources}}.

\subsubsection{Creating Benign Versions of Target Libraries.} 
\label{sec:evluation-creating-replacements}
For each of the three libraries, we considered associated source code and documentation to create equivalent benign versions. 
% We briefly describe them below.

\textbf{FingerprintJS} (v3 and v4), when loaded as a library from its \texttt{npm} package, provides a function, \texttt{load()}, which returns a Promise that resolves to an object that can be used to \texttt{get()} a unique \texttt{visitorId} for the user. While the computation includes privacy-harming code, the benign replacement that we developed only returns a randomly generated \texttt{visitorId}. We used a similar approach to the existing replacement included within uBlock origin\footnote{\url{https://github.com/gorhill/uBlock/blob/9123563895f0499849b4d85c4f95e1ed6ace2231/src/web_accessible_resources/fingerprint3.js}}, but modified the same to a Node.js module. FingerprintJS v2 has developers \texttt{get()} a unique \texttt{visitorId} without first needing to \texttt{load()} the library. Our benign replacement exports values for all versions of FingerprintJS without needing to individually support 52 minor versions released across 3 major versions (v2, v3, and v4). Considering the small number of export replacements, manually creating a replacement for FingerprintJS only takes around 20 minutes.

\textbf{Sentry} provides a \texttt{browser} library (\texttt{@sentry/browser}).
% as an entry point \texttt{npm} package~\cite{Sentry:NPM} to include its functionality within web applications. 
Below, we detail the development of replacements for the library.
\begin{enumerate}[leftmargin=*,nosep,noitemsep]
    \item \textit{Enumerating Replacements.} Sentry's browser library further imports related packages, which include utilities for tracing and profiling user action events and functionality to handle network requests, along with 24 functions (with \texttt{void} return values) from its core analytics library (\texttt{@sentry/core}). We used the official documentation as a guideline. Enumerating all imports, functions, and objects that needed replacements took us around 90 minutes of wall clock time.
    \item \textit{Functions.} We used a benign function that performed no operations (\texttt{noop}) as a stand-in for functions with no return values. We used the functions for the 24 functions imported from the core library, 3 functions imported to handle network requests, and one function each to support its additional \textit{``hub''} extension and to initialize the library. The library additionally exposes a function, \texttt{close()} (used to flush pending events), that returns a Promise that resolves with a timeout. We retained this function as in the original library. Once listed, since all functions required straightforward replacements, this took around 20 minutes. 
    % to add benign versions of functions.
    \item \textit{Class Definition.} Sentry defines a \texttt{BrowserClient} class with a constructor that accepts any argument it is passed. Developing a replacement for this class only takes a couple of minutes.
    \item \textit{Global Variable.} Finally, Sentry exposes all its functionality by defining an object instance of its \texttt{BrowserClient} class and exposes all aforementioned functions as members of this class. The library makes the object available as a global variable (on the \texttt{window} object). Also, it exports the same functionality to the user's code that imports it as an \texttt{npm} package. Defining the global variable and the \texttt{exports} takes under 5 minutes.
\end{enumerate}

Updates between Sentry versions (v5-v8) reuse exports of the same name, ensuring that, unlike for FingerprintJS, our benign replacements did not need to accommodate additional changes. As a result, the benign replacement that we developed in around 2 hours by wall clock time replaces 335 minor versions of Sentry's \texttt{npm} package.

\textbf{Prebid} provides a similar \texttt{npm} package to Sentry, i.e., its main functionality is included from an entry point module, while it imports additional functionality from related libraries. Our replacement focuses on the entry point module, i.e., \texttt{pbjs}, which, in turn, accesses other related functionality. The developer would need to refer to prebid's \textit{publisher}-facing APIs\footnote{\url{https://docs.prebid.org/dev-docs/publisher-api-reference.html}} which exposes 45 exports that we individually replace. We developed 41 \textit{function} replacements similar to those created for functions within Sentry (described earlier). Of these, 18 functions (\texttt{$\wedge$get$*$()}) returned an object, which we replaced with an empty object response. The remaining 23 functions did not have a return value, which we replaced with an empty, \texttt{noop} function. We replaced the remaining 4 exports with empty arrays that developers can define and edit.

While Prebid's library itself has undergone numerous changes across versions (\eg support for adapters, consent management, and user ID modules), its publisher-facing API has retained the same exports. We verified this against the library's release notes\footnote{\url{https://docs.prebid.org/dev-docs/pb9-notes.html}}
% \footnote{\url{https://docs.prebid.org/dev-docs/pb8-notes.html}}
% \footnote{\url{https://docs.prebid.org/dev-docs/pb7-notes.html}} 
and the library's source code. As a result, developing benign replacements for 41 functions and 4 arrays took only about an hour by wall clock time. Crucially, our replacement works against 186 minor library versions across 4 major Prebid version releases (v5-v9).

\subsubsection{Limiting Breakage.} We designed our replacements in a manner that would cause little breakage. Our replacements are expertly developed to maintain compatibility (uBO uses a similar approach for non-bundle replacements, e.g.FPJS\footnote{\url{https://github.com/gorhill/uBlock/blob/9123563895f0499849b4d85c4f95e1ed6ace2231/src/web_accessible_resources/fingerprint3.js}}). For each replacement module, we created a barebones application that accesses the library functionality. The barebones application helps evaluate each replacement, ensuring that the application remains functional. We package the resulting application and gather the webpack module corresponding to the benign replacement for use in \ the system's deployment. We evaluated our replacements by (i) ensuring that the JS execution of mock apps that bundle the library and the replacement were both successfully executed, (ii) our deployment returned metrics showing successful page load and DOM interaction events (Smith et al.~\cite{smith2022blocked} found network behavior highly predictive of breakage. Nisenoff et al.~\cite{nisenoff23} found users consider page load and responsiveness the most prominent breakage category.) We ensured our replacements have the shape (scope and exports) we want and are effective (execution of mock apps, page load, DOM responsiveness in the real world) against extensive testing. We can add a manual breakage evaluation by sampling modified websites.

% For each replacement module, we created a barebones application that accesses the library functionality. The barebones application helps evaluate each replacement, ensuring that the application remains functional. We package the resulting application and gathered the webpack module corresponding to the benign replacement for use in \system's deployment.

\subsubsection{Manual Breakage Analysis.} Core to understanding whether a website is \textit{functional} or \textit{broken} is the user's \textit{subjective perspective} of the features that are needed within the context of the website. With this in mind, we adopted the methodology presented by Snyder et al.~\cite{snyder_vibrate_2017} to study the effect of \system's replacements on website functionality. We sampled 30 websites (10 sites per target library) that bundled target libraries and on which are extension deployed benign replacements (determined from the crawl in \S\ref{sec:10k-crawl}). We instructed three undergraduate workers to visit each website twice in a row. In the first visit, they were instructed to use the website in an unmodified Firefox browser. Considering that the average dwell time for a user on a website is under a minute (Chao et al.~\cite{chao-2010}), the workers were instructed to perform as many actions on the page as possible in one minute. On the second visit, they were instructed to visit and use the same website within a Firefox browser with the extension enabled. We asked them to assign a score for each website: \textbf{1} if there was no perceptible difference between the two visits, \textbf{2} if the browsing experience was altered during the second visit, but they were still able to complete the same tasks as during the first visit, and \textbf{3} if they were unable to complete the same tasks as during the first visit.

We considered a website to be broken if the workers could not complete their intended task (i.e., the website was assigned a score of 3). We had the workers browse the same sites and record their scores independently and also had them include notes to provide context for their scores. We observed a high agreement ratio (96.67\%) between the workers' scores, who reported that 29 of the 30 websites were functional with benign replacements. For the single site that the workers reported broken (\texttt{\url{https://mlb.com}}), we visited the site with a modified version of our extension that does not perform replacement. We observed that the site was broken even without the replacements, primarily owing to longer wait times in loading large scripts as fragments and the extensions' \textit{blocking} network requests (see \S\ref{sec:performance-pipeline}). Our observations were also in line with the workers' notes, which indicated that they considered the page broken because it takes \textit{``too long to load''}. We, therefore, concluded that the site was broken due to the extension's network interception and not the benign replacements.

We note that we did not collect or process private data or identifiable information from our evaluators, and the
testing does not qualify as human subject research. Therefore, we did not seek an IRB review. This is consistent with prior work that used roughly the same evaluation setup~\cite{AdGraph,snyder_vibrate_2017,fingerprinting-fingerprinters}.

\subsubsection{Automated Breakage Analysis.} We also performed an automated analysis by crawling websites from the Tranco Top-1K. We visited each site within a Firefox browser with and without deploying \system. We used a puppeteer script to capture all console errors and page errors that were observed during each visit. Visits with \system deployed observed an average of 5 errors ($\sigma=15$;$M=1$), while visits without \system observed a slightly higher but similar average of 5.66 errors ($\sigma=14$;$M=1$). We noted that visits to 84\% sites ($n=$714/846) with \system observed the same number of errors or fewer. Interestingly, we found more sites that produced fewer errors with deploying \system ($n=$152;18\%) than sites that produced fewer errors without \system ($n=$132;15\%). We further analyzed instances where deploying \system produced more errors and made two notable observations. First, a majority of these errors refer to (a) rejections in setting cookies, (b) CSP violations, and (c) Timeout errors (potentially resulting from waiting for network responses). Second, 63\% ($n=$84/132) of these instances involve a difference $<=$2 errors between the two visits. We note that our analysis is limited and is only one of many insights into page breakage. With that caveat, considering that a large majority of websites respond similarly with and without deploying \system (overall average, \#sites with $<=$ errors, \#sites with $>=2$ errors), our observations show that \system retains web functionality and compatibility when deployed in the wild.

%% file: 05-discussion.tex
%-------------------------------------------------------------------------------
\section{Discussion}
\label{sec:discussion}
%-------------------------------------------------------------------------------

\paragraph{\textbf{Applicability to Other Systems}}
In this work, we target and evaluate against \js bundles generated by Webpack. We selected Webpack because it is the most popular bundling system on the Web. However, many other bundling approaches are used on the Web, some variations on the same approaches used in Webpack, some fundamentally different, and even working on different levels of the deployment process. We here briefly discuss these other bundling formats and how \system{} could be extended to apply to them (and with what difficulties).

Webpack is generally used to combine multiple different \js libraries and code units and to process them into a single \js file to make deployment easier and (in some cases) execution faster. While Webpack is the most common tool for this purpose, many others are used with different languages, build chains, testing frameworks, and, in some cases, bundling additional resource types beyond \js files. Examples of these alternative bundlers include Browserify~\footnote{\url{https://browserify.org/}} and Gulp~\footnote{\url{https://gulpjs.com/}}, among many others.

\system{} could easily be extended to cover these other bundling tools, as at root they all operate in the same manner (\ie{} consume multiple \js files, preprocess them, and then generate code for the resulting combined AST). To do so would only require generating new signatures for ASTs of each target privacy-harming library generated by the bundler's preprocessing and rewriting phases and understanding the structure of each bundler combines the ASTs generated by each input library into the final, resulting code unit. This is work that would only need to be done once per bundler version and then could be shared across all \system{} clients.

Another approach to \js bundling is directly combining each \js code unit into a single archive and shipping the entire archive to the client alongside the website's initial HTML. This approach, exemplified by Google's WebBundles~\footnote{\url{https://wpack-wg.github.io/bundled-responses/draft-ietf-wpack-bundled-responses.html}} proposal, does not preprocess or otherwise modify the include \js code units; included files are directly copied into the bundle archive. Cloudflare's  Cloudflare's Managed Components~\footnote{\url{https://managedcomponents.dev/}} product can also be seen as a form of this kind of bundling, though instead of all combined into a single archive, they're instead delivered ``on demand'', as managed by an overriding ``manager'' application, making any URLs unpredictable. Extending \system{} to cover these kinds of bundled applications would be trivial since bundled \js files are not modified in the bundling process. Identifying them within the bundled application is straightforward. Similarly, rewriting unwanted code is trivial since it only requires swapping the original file with the privacy-preserving alternative (either in the original archive or returned as a new subresource by the ``manager'' application).

% \system helps prepare content blocking mechanisms update their approaches for incoming and growing changes in the web. As privacy-harming actors adapt to a post third-party cookie world and gather new ways to circumvent blockers, approaches like \system counter such circumvention. While our analysis focused on webpack and example \texttt{npm} packages, the framework we present can be adapted to other bundlers and target libraries. For example, Browserify, another popular webpack-like bundler also adopts a modular approach that wraps each module with functions. However, each entry in its array of modules includes the wrapped function, an array of dependencies, and an array of exports. A similar approach can be developed to gather browserify modules and detect target libraries.

\paragraph{\textbf{Alternative Deployment Strategies}}
\label{sec:alternative-strategies}
In this work, we implement \system{} as a Firefox extension. We choose a Firefox extension because only Firefox's extension API includes the ability for an extension to buffer and rewrite a fetched subresource~\footnote{\url{https://developer.mozilla.org/en-US/docs/Mozilla/Add-ons/WebExtensions/API/webRequest/filterResponseData}} (like a \js file), before the file is seen, parsed, and executed by the \js engine.

However, the same approach we take in the extension could be identically deployed from other decision points, either to be more general across browsers (\eg{} as part of a man-in-the-middle proxy) or more specific to particular browsers (\eg{} as a modification to the V8 parsing pipeline). The majority of the resources needed for \system{} to work could apply equally across all of these intervention points (\eg{} fingerprints of the ASTs of unwanted to code, bundle ``unbundling'' logic, privacy-preserving replacement AST subtrees).

However, each of these intervention points would come with their own largely-predictable tradeoffs. Analyzing, unbundling, and rewriting \js bundles as part of a man-in-the-middle proxy would work for any browser or network tool, though with a performance impact (since \js engine optimizations like partial or lazy compilation would not be possible). Likewise, pushing \system{} logic into the \js engine directly would likely allow for greater performance at the cost of development and maintenance cost, and require browser-specific implementations. That said, we reemphasize that most of the novel aspects of \system{} would be generic and shareable across all possible intervention points.

% This example is dependent on network conditions and can affect page load times on bundles that contain large render-blocking code. Additionally, as of this writing Chromium-based browsers will soon be restricted in their use of a similar extension owing to manifest v3 restrictions. An alternative deployment strategy can introduce the \system intervention at function-granularity at runtime. This deployment would involve a V8 patch that analyzes incoming functions before choosing to execute them. Note that V8 parses code and generates an AST representation for that we can leverage to compare against known AST representations of tracking code. 

\paragraph{\textbf{Diversity of Target Library Representation}}
\system{} requires a significant amount of precomputation to work effectively. Specifically, \system{} requires precomputing a signature for every AST for each library or code unit that should be rewritten at runtime. While this is a significant improvement over the existing state of the art (\cite{SugarCoat:2021}, for example, requires the precomputing each target \emph{bundled application}, which will both be orders of magnitude larger in occurrence, but also enormously larger in terms of required disk space), it is still not trivial. The same target library can give different signatures depending on library version (\eg{} FingerprintJS v2 vs v3), bundler version (\eg{} Webpack v4 vs Webpack v3), bundler optimization strategies (\eg{} ``tree-shaking'' or no), library integration method (\eg CommonJS vs ECMAScript modules) among other dimensions.

As discussed in Section \ref{sec:framework}, \system{} uses several heuristics to effectively generalize signatures, to flatten the number of dimensions of signatures needed per target library (\eg{} label stripping, AST simplification, etc.). Nevertheless, there is still a tradeoff between coverage---generating as many signatures as possible to correctly identify the same target code across a wide range of representations---and concision---minimizing the memory, matching time, and disk space used on each \system{} client at runtime.

\paragraph{\textbf{Replacements}}
While other phases of \system are programmatically generated, we manually created benign replacements. Even with manual curation, we show in Section~\ref{sec:evluation-creating-replacements} that while benign replacements can take between 20 and 120 minutes to create, they can be scaled to replace a large number of versions of a target library (\eg Sentry's replacement applies to $>$300 versions). 

Regardless, this phase requires an expert understanding of both the specific bundler and the target library itself. While prior work has shown approaches that can automate this creation, we leave the adaption of a similar approach in the context of bundles as an avenue to explore in future work. Prior work like \cite{SugarCoat:2021}, which allows for the automatic creation of privacy-and-compatibility preserving versions of \js libraries, could be leveraged to greatly expand the number of target libraries \system{} can identify and rewrite in bundled applications.

\paragraph{\textbf{Caching Frequently Encountered Bundles}}
\label{sec:caching}
We designed \system to work against an extensive array of configuration options applied to bundles in the wild. Our deployment of \system makes an essential tradeoff: its large database of target library signatures consumes notable memory space (2.27 GB - 3.55 GB). In Section \ref{sec:database-size}, we discussed that reducing the database size can affect \system's detection of target libraries. However, developers can alternatively cache entire bundles that they frequently encounter. This way, while the browser extension would still block network requests, scripts do not need to be processed through \system on repeated encounters. For example, the service \texttt{\url{shopee.com}} has multiple related sites (\texttt{\url{shopee.com}}, \texttt{\url{shopee.co.id}}, \texttt{\url{shopee.co.th}}, \texttt{\url{shopee.tw}}), all of which load the same bundle (containing FPJS as one of the modules) from different URLs which would bypass filter lists. However, since the script's content is identical, a cache of this script can help perform detection and replacement without needing to process the script into component modules.

%% file: 06-relatedwork.tex
%-------------------------------------------------------------------------------
\section{Related Work}
\label{sec:related-work}
%-------------------------------------------------------------------------------

% This work contributes to and builds on an enormous body of research relevant to web privacy and content blocking.
% In this section, we highlight existing research that relates to our framework's design.

\paragraph{\textbf{Filter Lists.}} This research is closely tied to a broad domain of research investigating the advantages, effectiveness, and responses to filter list-based content blocking. Note that existing filter lists (e.g., EasyList, EasyPrivacy) are developed manually and community-maintained. Merzdovnik et al.~\cite{merzdovnik2017block} performed a large-scale study that highlighted the effectiveness of filter list-based browser extensions and also indicated that these tools often lead to a \textit{decrease} in overall CPU usage, even when considering their own overhead. Gervais et al.~\cite{gervais2017quantifying} quantified the privacy gain from ad-blockers and discovered that these tools can reduce interactions with third-party entities by up to 40\% with default settings.
However, other studies have pointed out significant inefficiencies in filter lists. Snyder et al.~\cite{Snyder2020Sigmetrics} highlighted the abundance of ``dead-weight'' rules that offer no discernible benefits in popular lists. Similarly, Alrizah et al.~\cite{Alrizah2019IMC} found that popular lists contain a large number of false positives, which can take two or three months to be discovered.

\paragraph{\textbf{Automated Content Blocking Approaches.}} Numerous studies have developed approaches to either help automatically generate filter list rules or develop alternative methods to block privacy-harming content. AdGraph~\cite{AdGraph} created a graph-based machine learning approach that used features like URL length and origin to differentiate between benign and privacy-harming resources. Bhagavatula et al.~\cite{Bhagavatula2014} also used URL-based features to train a machine-learning model for resource filtering. Chen et al.~\cite{Chen2021EventLoop} used filter lists as ground truth and developed behavioral signatures based on the \js event loop, while Sun et al.~\cite{Sun2021Infocom} classified \js execution based on Web API calls. Le et al.~\cite{AutoFR} developed a reinforcement learning framework that generates filter list rules specific to a site of interest. They showed that their approach was comparable in visual breakage to manually-created filter lists.

\paragraph{\textbf{\js Analysis.}} Prior work that performs static or dynamic analysis on \js to create ``pre-filters''~\cite{Fass2019JSTap} for malicious scripts, detect malicious scripts that camouflage as benign scripts~\cite{Fass2019HideNoSeek}, and detect scripts that evade detection by adopting various obfuscation strategies~\cite{Fass2018Jast}. In their analysis of the Top 10K sites in the Alexa list, Moog et al.~\cite{Moog2021DSN} found that 90\% sites contained a minified or obfuscated script. Fouquet et al.~\cite{jsripper} presented a static analysis approach to webpack bundles but created signatures based on serializing their AST and hashing the result (compared to the complex, Merkle tree-based approach) we present. 
% They incorrectly assume that \textit{``...it is unlikely in real-world scenarios that the bundling and minification processes would modify the AST of functions beyond renaming local bindings.''} 
Their work does not consider the effect of various bundling and minification options, which our work highlights as highly importance in ensuring the effectiveness of \system as a real-world deployment. 
Amjad et al.~\cite{amjad_blocking_2023} employed stack traces (using dynamic analysis) to identify functions responsible for tracking within scripts. Their work addresses tracking in \js but is not robust against bundles, which include nested functions and commonly introduce code transformations and stack information changes. Importantly, their proposal addresses \textit{``tracking behavior''}, while \system identified specific libraries. Since our \system is not similarly limited in its ability to identify specific libraries, the benefits provided by the two approaches can be combined. \system can be incorporated as a first step to help narrow the scope of privacy-harming code to a small part of a large, bundled resource. Thereafter, ~\cite{amjad_blocking_2023}'s tool can identify specific functions within the library that perform privacy-harming actions, further increasing the granularity of detection and attribution. 
% Further, the replacements developed for URR can be modified to benefit attribution returned from Amjad et al.'s solution. 

More relevant to our work, Rack and Staicu~\cite{Rack2023CCS} presented a method for detecting and partially reverse engineering bundles. Importantly,~\cite{Rack2023CCS} overlaps with the \textit{Gather Modules} phase (\S~\ref{sec:gathering-modules}) for which we developed our own approach. To elaborate on the differences between the two approaches, we must divide the \textit{Gathering Modules} phase into two steps. First,~\cite{Rack2023CCS} determines if a script is a bundle by checking for specific text.
%  (allowing them to detect 12 configurations).
% \footnote{\url{https://github.com/zenoj/BundlerStudy/blob/main/Fingerprints/webpack.js}}).
Their use of keywords and consideration of a small set of configurations will have limited effect against bundle configurations in the wild. Our approach instead considers \textit{code structure} and can handle code transformations that would frustrate~\cite{Rack2023CCS}. Finally, they did not evaluate their approach's precision, recall, or accuracy against scripts gathered in the wild (see Table~\ref{tab:bundle-sampling}).
Second, after identifying bundles,~\cite{Rack2023CCS} relies on source maps and unmodified module names to parse bundles and break them down into component modules, restricting their approach to only work on 10\% of real-world bundles. We instead present an automated approach (see \S~\ref{sec:processing-modules} and \S~\ref{sec:comparing-modules}) that can reverse engineer bundles regardless of the accidental availability of source maps. Our approach does not rely on either of these developer choices and works on a broad range of real-world sites.

%% file: 07-conclusions.tex
%-------------------------------------------------------------------------------
\section{Conclusion}
\label{sec:conclusion}
%-------------------------------------------------------------------------------

Content blocking plays a crucial role in safeguarding privacy, enhancing performance, and preserving user autonomy online. However, the increasing use of bundlers presents a challenge -- websites often intermingle tracking code with benign code within a single script, rendering traditional URL-based content blockers ineffective.
We present a framework, \systemLong (\system), that detects bundles and reverse engineers them back to constituent modules. We develop an approach to identify the privacy-harming modules that we replace with benign alternatives. We demonstrate the effectiveness of our system in identifying Webpack bundles and further develop signatures for a fingerprinting library (FingerprintJS), an advertising library (Prebid), and an analytics library (Sentry). Our implementation can identify the bundled versions of these libraries in the wild via a similarity threshold that minimizes false negatives and prevents false positives within our training data. Leveraging our approach, we found the use of these libraries within bundled scripts on 697 sites of the Tranco 10K.
% -- uses which current content blocking approaches cannot successfully counteract\kaytwo{please confirm I am not overstating here}. 
Further, we implemented a prototype deployment of \system as a browser extension and observed that the extension only adds 0.5s to the DOM Interaction Time on a page visit. \system can expand existing content-blocking approaches to combat the use of bundlers to hide privacy-harming code. It can be further adapted to combat alternative bundling strategies and can be further used to detect the use of other privacy-harming libraries to bolster web privacy protections. 